\documentclass{aa}

\usepackage{graphicx}
\usepackage{txfonts}
\usepackage{natbib}
\usepackage{subfig}
\usepackage{graphicx}
\usepackage{natbib}
\usepackage{hyperref}

\usepackage[scientific-notation=true]{siunitx}

\newcommand{\HII}{\mbox{H\,{\sc ii}}}
\begin{document}
\titlerunning{Cloud
fragmentation under rotation and gravity}

\title{Revealing a spiral-shaped molecular cloud in our galaxy -- Cloud
fragmentation under rotation and gravity }
%\subtitle{How cloud fragments and produces star clusters}
\author{Guang-Xing Li \inst{1, 2}  \and Friedrich Wyrowski\inst{2} \and
Karl Menten
\inst{2}}
 \institute{$^1$University Observatory Munich, Scheinerstr. 1, D-81679
 M\"unchen, Germany, $^2$Max-Planck-Institut f\"ur Radioastronomie, Auf dem
 H\"ugel, 69,  53121 Bonn, Germany}
\offprints{Guang-Xing Li, \email{gxli@usm.lmu.de}}

\abstract{The dynamical processes that control star formation in
molecular clouds are not well understood, and in particular, it is unclear if
rotation plays a major role in cloud evolution.
We investigate the importance of rotation in cloud
evolution by studying the kinematic structure of a spiral-shaped
Galactic molecular cloud G052.24$+$00.74. 
The cloud belongs to a large
filament, and is stretching over $\sim 100 \;\rm pc$ above the Galactic disk
midplane. The
spiral-shaped morphology of the cloud suggests that the cloud is
rotating.
We have analysed the kinematic structure of the cloud, and study the fragmentation and
star formation. 
We find that the cloud exhibits a regular velocity
pattern along west-east direction -- a velocity shift of
$\sim 10\;\rm km\;s^{-1}$ at a scale of $\sim 30$ pc. The kinematic structure of
the cloud can be reasonably explained by a model that assumes rotational
support. Similarly to our Galaxy, the cloud rotates with a prograde motion.
We use the
formalism of Toomre (1964) to study the cloud's stability, and find that it
is unstable and should fragment.
The separation of clumps can be consistently reproduced assuming
gravitational instability, suggesting that fragmentation is determined by
the interplay between rotation and gravity.
Star formation occurs in massive, gravitational bound clumps. Our analysis
provides a first example in which the fragmentation of a cloud is regulated by the
interplay between rotation and gravity. }

% \date{\today}
\maketitle

%opening

\section{Introduction}
Star formation takes place in the dense and shielded parts of the molecular
interstellar medium. An increasingly dynamical picture of cloud evolution
has been revealed by recent observations and simulations \citep[][and
references therein]{2014prpl.conf....3D,2015ARA&A..53..583H}.
Star formation may be determined by a combination of turbulence
\citep{2004RvMP...76..125M,2012A&ARv..20...55H}, gravity \citep{2009ApJ...699.1092H,2011MNRAS.411...65B,2012MNRAS.427.2562B},
magnetic field \citep{2014prpl.conf..101L} and ionisation radiation
\citep{1994A&A...290..421W,2009MNRAS.398.1537D}.

The hierarchical structures of molecular clouds are produced by a series of
fragmentation processes. In the theory of turbulence-regulated star formation
\citep{1999ApJ...527..285B,1999ApJ...526..279P,2004RvMP...76..125M,2005ApJ...630..250K},
supersonic turbulence creates a set of density fluctuations, and it is the
high-density parts that undergo gravitational collapse. It has also been
alternatively proposed that the evolution of molecular clouds is governed by gravity, and gravitational collapse
creates a hierarchy of structures which then form stars
\citep{1953ApJ...118..513H,1973ARA&A..11..219L,1984MNRAS.210...43Z,2008ApJ...689..290H,2011MNRAS.411...65B}.
It is unclear what dominates the evolution of molecular clouds. 

% The
% importance of angular momentum in star-forming clouds has also been
% discussed \citep{1995ARA&A..33..199B}. However, it is unclear how important it
% is in star-forming regions.

In recent years, the importance of environment on cloud evolution has been
addressed. It is now relatively well-recognised
that molecular clouds are not isolated objects. They can
belong to large-scale structures
\citep{2014A&A...568A..73R,2015MNRAS.450.4043W,2015arXiv150608807Z,2016A&A...591A...5L,2016A&A...590A.131A,2013A&A...559A..34L},
suggesting a connection between Galactic shear and cloud evolution. A connection
between cloud evolution and large-scale magnetic field has also been suggested \citep{2015Natur.520..518L}.
Moreover, cloud evolution can be significantly influenced by stellar
feedback \citep{1977ApJ...214..725E,1994A&A...290..421W,2002MNRAS.329..641W}. All these
effects have been proven to be important at least in some cases.  Nevertheless,
the role of angular momentum in molecular clouds is still unclear.

In this work, we present a study of a spiral-shaped molecular cloud
G052.24$+$00.74. The cloud is identified from the Galactic Ring Survey
\citep{2006ApJS..163..145J}. It belongs to a large ($\sim$ 500 pc) gas
filament (filamentary gas wisp) discussed in an earlier work
\citep{2013A&A...559A..34L}. The molecular gas of the cloud is distributed in spiral-arm-like
features, and we have named it ``Spiral Cloud''.
The whole cloud exhibits a regular velocity pattern  and clear signs of
fragmentation on the ``spiral arm'' part of the cloud.  A star cluster has already formed in
one of the clumps. In this paper we focus on the 
global kinematic structure of the cloud, and addresses its connection with the
ongoing star formation activities.

\section{Archival data}

We used $^{13}$CO(1-0) molecular line data ($\nu_0=110.2\rm\; GHz$) from
the Galactic Ring Survey \citep{2006ApJS..163..145J}, which is a survey of the Milky
Way disk with the SEQUOIA multipixel array on the Five College Radio Astronomy
Observatory 14 m telescope, and covers a longitude range of
$18^{\circ}<l<55.7^{\circ}$ and a latitude range of $|b| < 1^{\circ}$ with a
spatial resolution of $46\arcsec$. The beam efficiency is  $\eta_{\rm mb}$ = 0.48
\citep{2006AJ....131.2921R}. At a distance of 9.8 kpc
\citep{2013A&A...559A..34L}, the spatial resolution is around 2.2 pc. The
velocity resolution is $0.22\;\rm km/s$. For our region, the rms sensitivity is
$\sigma(T_{\rm A^*}) \approx 0.24\,\rm K$. These observations are
velocity-resolved which allowed us to trace and analyse the cloud's kinematic structure in detail.

We have made use of  3.6 $\mu$m and 8 $\mu$m data from the
GLIMPSE project \citep{2003PASP..115..953B}, which is a fully sampled,
confusion-limited, four-band near-to-mid infrared survey of the inner Galactic
disk. We use 24 $\mu$m data from the MIPSGAL project
\citep{2009PASP..121...76C}, which is a survey of the Galactic disk with the
MIPS instrument on {  Spitzer} at 24 $\mu$m and 70 $\mu$m. 
The 3.6 $\mu$m emission is sensitive to the presence of YSOs, and the 8 $\mu$m
emission typically traces polycyclic aromatic hydrocarbon (PAHs).
Star formation can be traced by the 24 $\mu$m emission, which originates from
the dust heated by newly-born stars.

% 
%  We use the $^{13}$CO(1-0) observation from the FCRAO telescope to trace the
% structure of the cloud. The observations are velocity-resolved, which enable us to
% analyse the kinematic structure of the cloud in details. 
% Complimentary data from the {  Spitzer} telescope are obtained.

\section{Results}
\subsection{Multi-scale structure}
\begin{figure*}
\includegraphics[width=1 \textwidth]{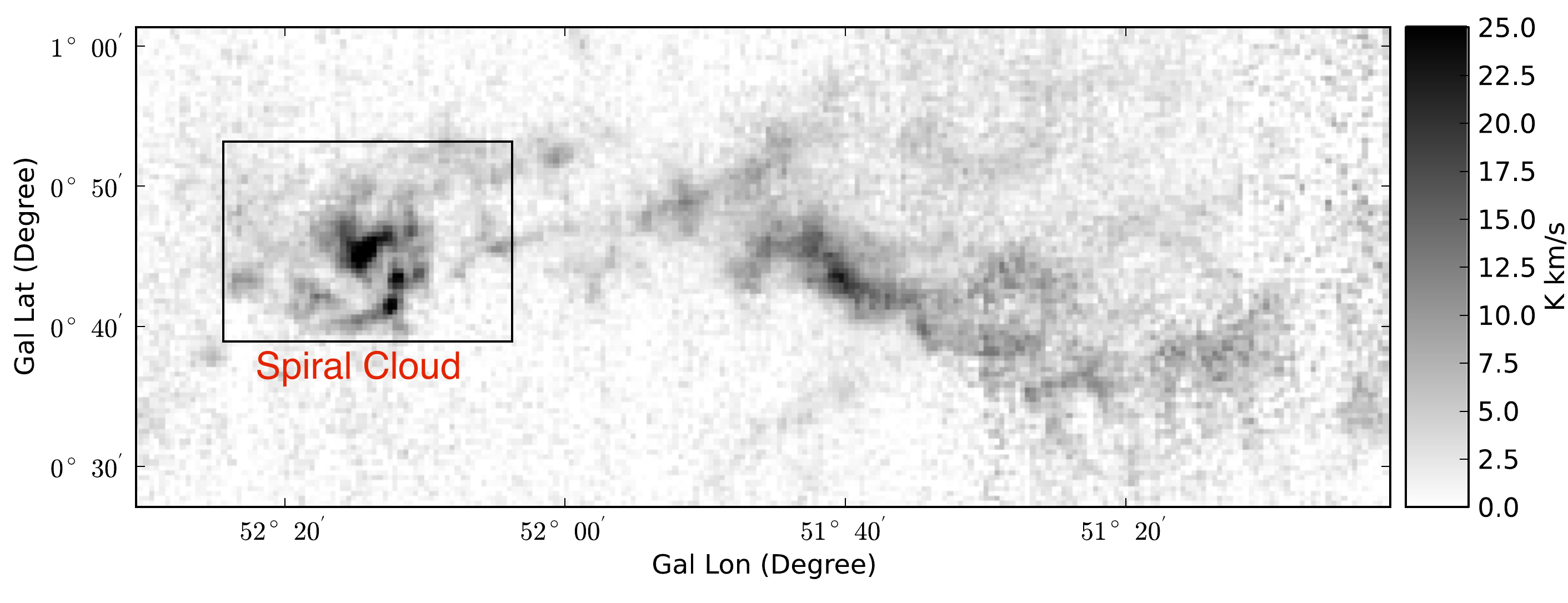}\\
\includegraphics[width=0.53 \textwidth]{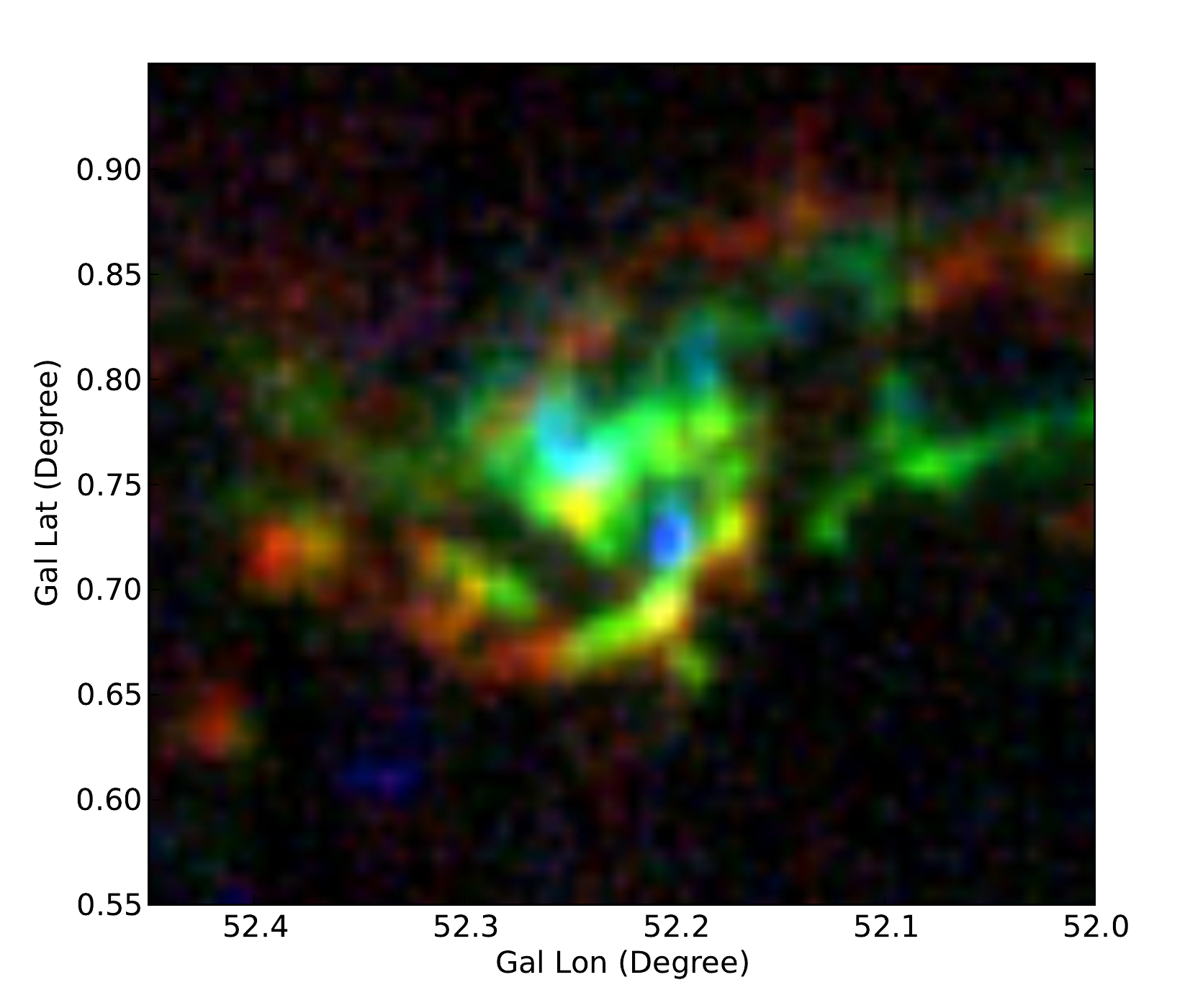}
\includegraphics[width=0.5 \textwidth]{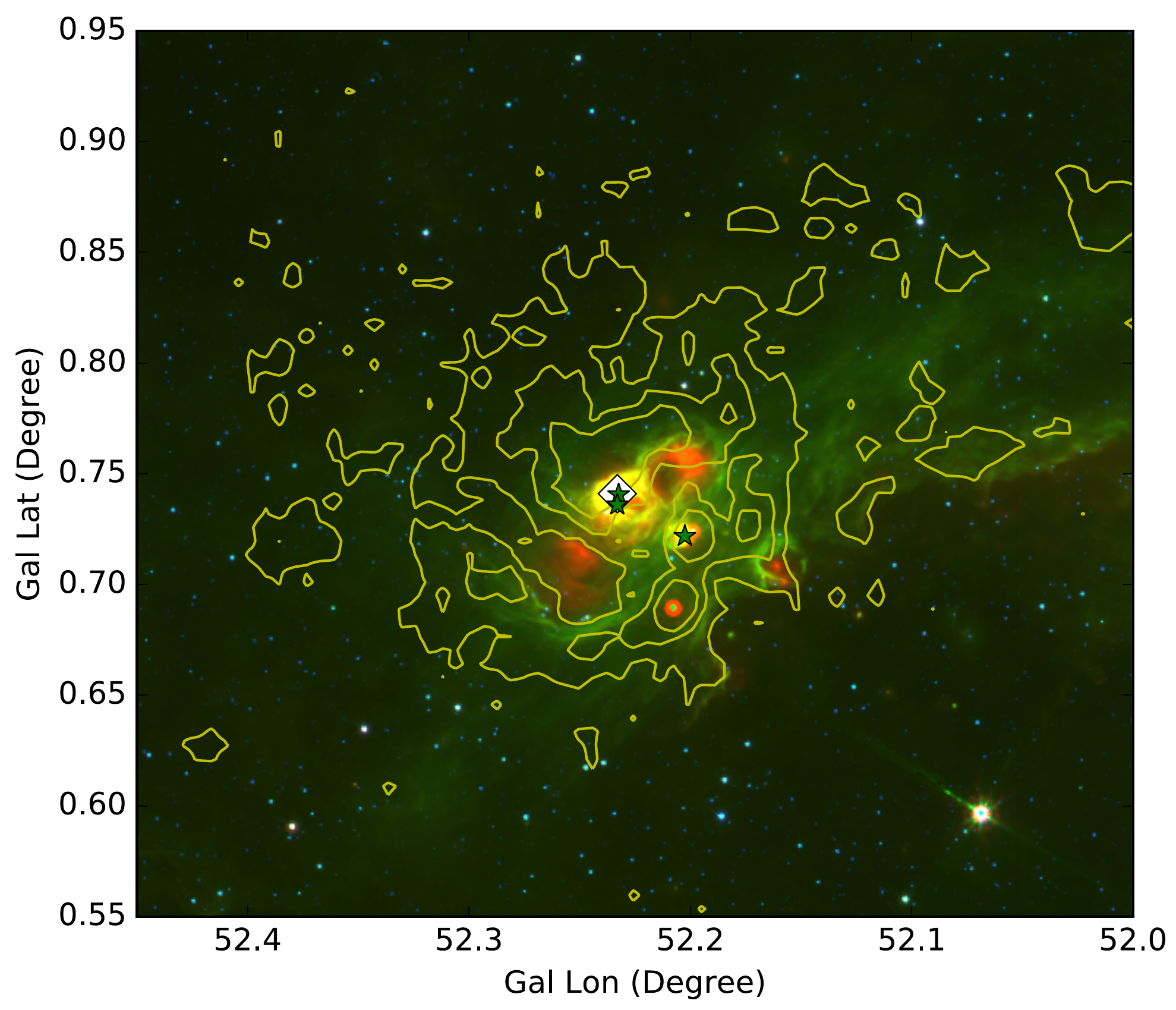}
\caption{{  Upper panel:}  $^{13}$CO(1-0) emission integrated from
-4.5 to 15.0 $\rm km\;s^{-1}$ from the GRS survey \citep{2006ApJS..163..145J}.
The cloud G052.24+00.74 is on the left side. This cloud is connected with the
neighbouring cloud G051.69+00.74 by some filamentary gas wisps.
Channel maps of the region can be found in Appendix \ref{appendix:B}.
{  Lower left panel:} Composite three-color image of the $^{13}$CO(1-0)
emission from the Spiral Cloud G052.24+00.74. Red:
$6.52< \textit{v}_{\rm lsr}<15.0\; \rm km\; s^{-1}$
Green: $3.12< \textit{v}_{\rm lsr}<6.10 \; \rm km\; s^{-1}$
Blue: $-4.53< \textit{v}_{\rm lsr}<2.7\; \rm km\; s^{-1}$. 
{  The velocity centroid
map and velocity FWHM map can be found in Appendix \ref{sec:appen:v}}.
{  Lower right panel:}
Spitzer GLIMPSE \citep{2003PASP..115..953B}
and MPISGAL three-color image of the Spiral Cloud G052.24+00.74.
Red: 24 $\mu$m, Green: 8 $\mu$m, Red: 3.6 $\mu$m. Overlaid contours
are the velocity-integrated $^{13}$CO(1-0) emission {  (integrated from $-4.5
{\rm km/s}<  \textit{v}_{\rm lsr}< 11.5  \;\rm km/s$)}.
Contours correspond to {  5.1, 10.2, 15.3 $\rm K\; km\;s^{-1}$}.
The white diamond  at the centre stands for a star cluster discovered
in the GLIMPSE survey \citep{2005ApJ...635..560M}\label{fig:cloud52},
and the green stars stand for \HII$\ $ regions collected from the literature
\citep{1989ApJS...71..469L,2009A&A...501..539U}. }
\end{figure*}

From the $^{13}$CO data of the GRS survey (Fig. \ref{fig:cloud52}) we
found that the cloud belongs to a larger system of two clouds (the Spiral
Cloud, G025.24$+$00.74 and G051.69$+$00.74).
This double cloud system has a total mass of $1.2\times 10^5\; M_{\odot}$ \citep{2010ApJ...723..492R}
and a total physical extent of $\sim 140\; \rm pc$. Wisps of molecular gas connect the cloud
with another cloud at $l\sim 51^{\circ}$ (G051.69+00.74). 
The lower boundary of the two clouds is arc-like, and 
is associated with the G52L nebula in  \citet{2012ApJ...759...96B}.
% The two clouds are connected with some wispy gas filaments. 

This large double cloud system
stretches to a $\sim 500\;\rm pc$  {filamentary gas wisp}
\citep{2013A&A...559A..34L}. Recently, such filamentary
structures have been found to be relatively 
 common throughout the Galaxy
\citep{2014A&A...568A..73R,2015MNRAS.450.4043W,2015arXiv150608807Z,2016A&A...591A...5L,2016A&A...590A.131A}
and are predicted by theory and simulations
\citep{2014MNRAS.441.1628S,2015MNRAS.447.3390D,2001MNRAS.327..663P}.

%  Seen from $^{13}$CO
% emission from the GRS survey \citep{2006ApJS..163..145J}, the cloud exhibit a
% lopsided spiral-shaped morphology.
% The $^{13}$CO emission from the cloud also exhibit a regular velocity patten:
% gas at the south-east part tends to be red-shifted and gas at the north-wast part of
% the cloud tends to be blue-shifted. The cloud is fragmented and clumps can be
% found in the cloud.

\subsection{Structure of the cloud}
\label{sec:structure}
The whole Spiral Cloud has a mass of $2.72 \times 10^4 M_{\odot}$  and a
radius of $\sim 12\; \rm pc$ \citep{2010ApJ...723..492R}. It has
 a centrally-condensed morphology, and exhibits a regular velocity
pattern {  (Fig.
\ref{fig:cloud52}, \ref{fig:mom})} where the south-eastern part of the cloud is
red-shifted and the north-western part of the cloud is blue-shifted. {  The spiral-shaped
structure lead us to assume that the velocity
difference originates from rotation.
The inferred cloud rotation is similar to the Milky Way rotation,  and the
projected angle between the cloud rotational axis and the rotation axis of the Galactic disk is estimated to be 22.5$^{\circ}$ (measured directly
from the map).} The rotation is also visible in channel maps (Fig.
\ref{fig:channel5152}). {  Velocity centroid map and velocity FWHM map
can be found in Appendix
\ref{sec:appen:v}.}

\begin{figure*}
\includegraphics[width=\textwidth]{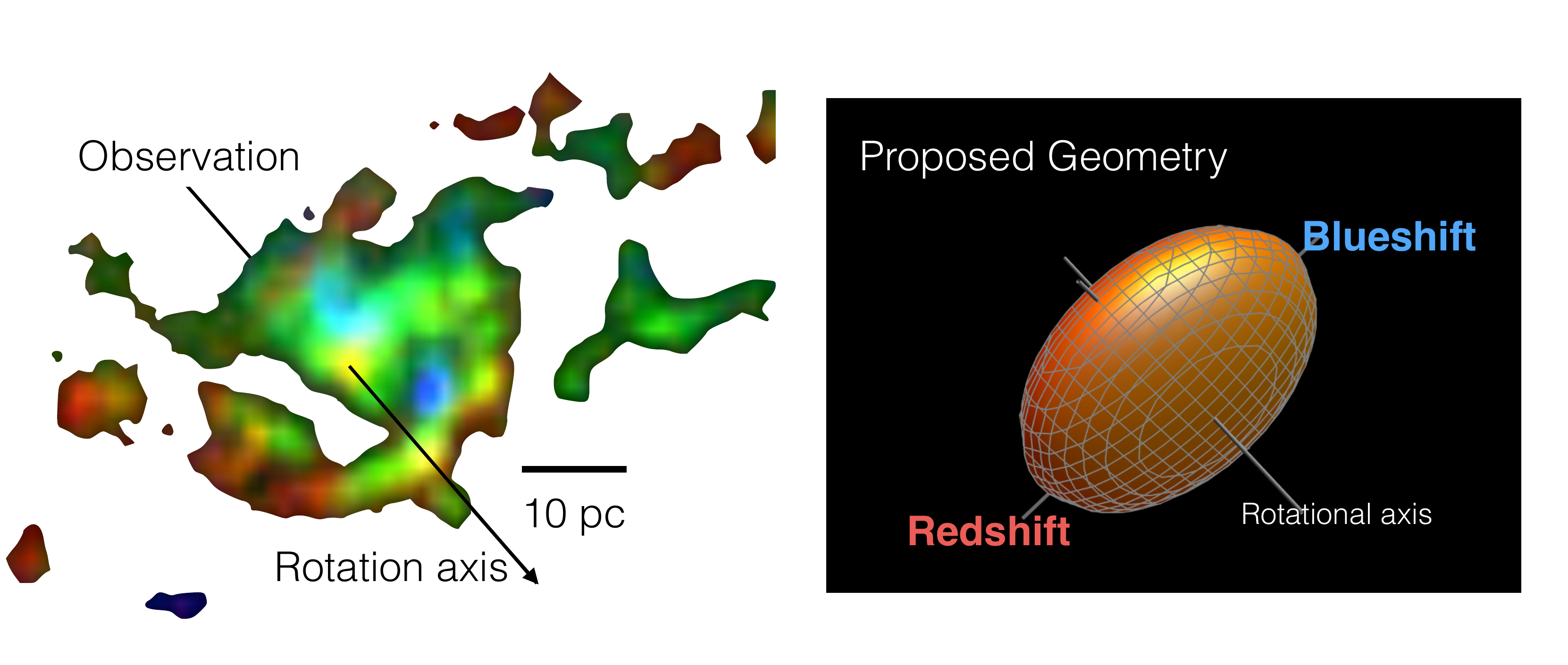}
\caption{  An illustration of the geometry of the Spiral Cloud. The left panel
is a 3-colour rendering of the spiral cloud. Red colour stand for redshifted gas,
and blue colour stand for blueshifted gas. On top of the image, we overlay
the suggested rotational axis.
In the right panel, we present a 3D ellipsoid, with two longer axes and one
shorter axis. The orientation mimic the suggested orientation of the spiral
cloud.
The ellipsoid (representing the Spiral Cloud) is rotating with respect to its
shorter axis, which we have indicated in the plot. The western side of the
Spiral Cloud is rotating toward the observer. The angle between the rotational axis of the Spiral Cloud and the
rotational axis of the Milky Way disk is estimated to be $22.5^{\circ}$, and the
cloud inclination is estimated to be $45^{\circ}$.
\label{fig:illustration} }
\end{figure*}
% A visual examination of the data suggests that we are viewing the
% % cloud with an inclination of $\sim 45^{\circ}$.

{  Because of the observed rotational pattern, we assume that in 3D the cloud
can be approximated as an ellipsoid characterised by two longer axis and one
shorter axis, and is rotating with respect to its shorter axis.} 
 A prior, the cloud's inclination is unknown \footnote{  Here,
 an inclination angle of 0$^{\circ}$ means we see the cloud face-on, and an
 inclination angle of 90$^{\circ}$ means we see the cloud edge-on.}, and from
 the map we estimate an inclination of 45$^{\circ}$. {  Because of projection,
 one can not readily tell which part of the cloud is
 closer to us in 3D. However, we can still infer the 3D geometry by assuming that the
 spiral arms are trailing, and thus we infer the orientation of the cloud as
 illustrated in Fig.
 \ref{fig:illustration}.}

  To better analyse the cloud structure from the centre to the outside, we
  divided the $^{13}$CO(1-0) map into rings. Ideally, each of the ring
  corresponds to a circle were the cloud is viewed face-on. 
% Practically, first we define the
% position where the column density estimated from the $^{13}$CO(1-0) reaches its maximum as the centre
% position.
{  We measured the longer and shorter axes manually from the map.}
 Based on this, we divided the cloud into different rings. {  The
semi-major axes of the rings (which correspond to the radii of the circles in 3D
when de-projected)} are called  {cloud radii}.
{  The widths of the rings were chosen to be 2.7 pc, which is larger and still comparable to the resolution
derived from the beam size (2.2 pc). }  The structure of the individual rings can be found in Appendix \ref{sec:appen:vg}.
The velocity structure of the cloud can be well seen in each ring.
%  There are,
% however, some portion of gas does not follow the {  expected}overall velocity
% structure.
% An inspection of data cube in the 3D position-position-velocity (PPV)  space reveals that these belongs to
% smaller ``cloudlets'' that {  probably belong to the whole complex, but might 
% not necessarily belong to the spiral structure} \footnote{One example of such
% structures is clump 11 in Fig. \ref{fig:leaves}. In general, these structures
% have significantly different velocities as compared to the bulk of the Spiral
% Cloud (See Table \ref{table:1})}.
% Since they do not appear to contribute much to the overall flux, we choose not
% to remove them when analysing the cloud structure. One {  such}
% source is clump ``0'' in Fig. \ref{fig:leaves} which has a mass of $\sim
% 10^{3}\; \rm M_{\odot}$. By inspecting Figs.
% \ref{fig:rings1} -- \ref{fig:rings3}, we estimate that this can lead to a slight
% over estimation of cloud mass to $\sim$ 10 \%.

For each ring, we measured the mean column density. 
The velocity range was chosen to be $-4.5 {\rm km/s}<  \textit{v}_{\rm lsr}< 11.5  \;\rm
km/s$ to cover the whole cloud. To focus on the spiral structure, we only
analysed regions that have $\int T_{\rm A*}\rm {dv} \geq  5.1 \rm
K\;km\;s^{-1}$.
This corresponds to 3.5 times the rms noise level.
The 
corresponding region is indicated in Fig. \ref{fig:rings}.

 We estimated the $\rm H_2$ column density using ${N}_{\rm H_2}
= 5 \times 10^{20}\;{\rm T}_{\rm mb}\, {\rm d} {\rm v} / (\rm K\; km\; s^{-1})$
\citep{2001ApJ...551..747S}. {  ${\rm T}_{\rm mb} = {\rm T}_{\rm A*} / \eta_{\rm
mb}$, where ${\rm T}_{\rm A*}$ is the antenna temperature and $\eta_{\rm mb}$ is
the beam efficiency, $\rm v$ is the velocity. } This conversion has been derived
assuming an excitation temperature ${ T }_{\rm ex} = 10 \rm K$, and a $^{13}$CO abundance ${\rm R}(^{13}{\rm CO/H_2}) = 1.7\times 10^{-6}$. Fig.
\ref{fig:cloud:structure} shows the mean surface density as a function of the
sizes of the rings (which we named  {Cloud Radius} $r_{\rm cloud}$). From Fig. \ref{fig:cloud:structure},
we found that mean column density measured in H$_2$ can be expressed as a function of cloud
radius $r_{\rm cloud}$:
\begin{equation}\label{eq:sigma:r}
N_{\rm H_2} =  1.5 \times 10^{22} {\rm cm^{-2}} \times \big( \frac{r_{\rm
cloud}}{\rm 5\; pc}\big )^{-1}\;.
\end{equation}
The surface density of the
cloud can be estimated as 
$\Sigma = N_{\rm H_2} \times m_{\rm H_2} \times 1.36$ where $m_{\rm H_2}$ is the mass of $\rm H_2$ and $1.36$ is the correction for
Helium and other heavy elements.
At the outer regions, Eq. \ref{eq:sigma:r} captures the structure of the cloud
quite well, but at the inner $\sim 3$ pc region, there are some noticeable
differences between the column density in the analytical model and the
observational data. However, this discrepancy is not severe  since this is
also the region where the $^{13}$CO is likely to be optically thick and the
observations are underestimating the column density.

Assuming that the gas in the cloud stays in a flattened disk, 
{  by taking advantage of the the fact that the column density scales as
$r_{\rm cloud}^{-1}$, one can analytically evaluate the expected rotation
profile following \citet{1963MNRAS.126..553M}. Disks that share this kind of
density profiles are called ``Mestel disks''.
The rotation velocity profiles can be determined from the balance between
centrifugal force and gravitational force, they are flat,
characterised by constant circular velocities. In the case of our disk,  the
model is accurate only at $r_{\rm cloud} \gtrsim 5 \;\rm pc$ where the model
does capture the density structure to a reasonable accuracy.} The theoretical 
circular velocity can be computed as
\begin{equation}\label{eq:v:virc}
{\rm v}_{\rm circ}^{\rm model} =  2\;\pi\; G\; \Sigma_0\; r_0 = 6.6\;\rm km/s,
\end{equation}
where $\Sigma_0  r_0$ can be found in Eq. \ref{eq:sigma:r}, $r_0 = 
5 \;\rm pc$ and $\Sigma_0$ can be evaluated from Eq. \ref{eq:sigma:r}.
Assuming that the cloud is rotationally supported, the
absolute velocity
difference between the left and right side of the disc is thus of $2 \rm
\; v_{\rm cir}$ (i.e.
13.2 $\rm km/s$).
% 
% 
% {  Given the density distribution, one would expect
% the system to have a constant velocity on one side and another constant velocity
% on the other side. The velocity difference is $2 \times v_{\rm circ}$. 
% % Ideally,
% % there should be a velocity jump at the very centre of the cloud, and because our
% % disk deviate from the Mestel disk at the very centre ($r_{\rm cloud} < 5
% % \;\rm pc$), we expect a somewhat smoother transition. 
% }

 Assuming an
inclination of $45^{\circ}$, the predicted velocity shift of either side of
the disk is 4.68 $\rm km/s$ with respect to the systemic velocity.
This is the theoretically-expected velocity shift if we assume that the
cloud is gravitationally bound and has a disk-like geometry. 
% 
% Another way to
% estimated the theoretically-expected velocity shift is to assuming virial
% balance.
% The velocity shift at a given radius can be evaluated as 
% \begin{equation}
% {\rm v}_{\rm circ}^{\rm virial} = \sqrt{5\; G m / r_{\rm cloud}} \times 2.35\;.
% \end{equation}
% where we can measure $m$ from the observational data, and 2.35 comes from the
% fact $\rm FWHM = 2.35 \; \sigma$.
% 
% Both ${\rm v}_{\rm circ}^{\rm model}$ and ${\rm v}_{\rm circ}^{\rm virial}$ give
% us an estimate of the velocity shift we would expect if the cloud were
% gravitationally bound. ${\rm v}_{\rm circ}^{\rm model}$ is more accurate since
% it takes the flattened geometry into account.

%  In Fig.
% \ref{fig:cloud:structure}, we compare the theoretical values with the observed
% velocity shifts at different cloud radii. {  The velocity shift is estimated
% from the mass-weighted velocity dispersion of gas in the individual rings, and
% we have used the formula\footnote{We have verified this equation numerically by
% creating such a ring and then evaluate its velocity dispersion.} $v_{\rm circ}^{\rm obs} = 1.4 \times \sigma_v /
% {\rm cos}(\theta) \approx 2 \times \sigma_{\rm v}$. Here we have assumed an
% inclination angle of 45$^{\circ}$.
% }
{ 
In Appendix \ref{sec:appen:vg} we compare the velocity structure of the cloud
with the expected velocity structure  derived from the
\citet{1963MNRAS.126..553M} model.}
% We find a reasonably good agreement between the theoretically-predicted
% velocity structure and the observations. 
In general, the agreement is better at $r_{\rm cloud} \gtrsim 15\;\rm pc$. Inside
the inner 15 pc, the velocity difference is not always obvious. However, in this
region, the model is not so accurate because of the deviation of the real
density profile from the density profile of the \citet{1963MNRAS.126..553M}
model, and the data are not accurate either, because of the observed line widths are typically large (around a few $\rm km/s$). 

The  agreement between the model and the data leads us to conclude
that the cloud is probably rotationally-supported.
We note, however, that there are still structures that do not follow this
regular rotation pattern. A few explanations are possible: first, the cloud is already
fragmented, and for each ring, only a small portion is sampled by the molecular
gas. Thus this incomplete sampling introduces some irregularities to our data.
Second, for each line of sight, 
a typical line width of $ \sim 1 \; \rm
km/s$ is common (see e.g. Table \ref{table:clumps}).
Third, the cloud is already fragmented, and the very process of gravitational
instability can introduce significant deviations from regular circular rotation.  The
gas motion would also be influenced by the expansion of the embedded H$_{\rm \sc
II}$ regions. These observational and theoretical
uncertainties can potentially account for the observed velocity irregularities.

% Even though the cloud have a regular velocity pattern, the velocity profile does
% not correspond exactly to what we would expect from a rotationally-supported
% disk.
% Some of the gas has significantly
% lower velocities. Contaminating gas structures are also identified in the
% data. However, such deviations are exactly what we expect if the cloud fragments
% under rotation and gravity.
% This is because if the cloud fragments, the very process of fragmentation will
% introduce disturbances to the overall velocity pattern.
% Molecular clouds are also dynamically interacting with their environments, and
% gas in the cloud does not have much time in isolation and thus can not
% relax to a completely regular pattern.

\subsection{Fragmentation}
We used  the \texttt{Dendrogram} program
\citep{2008ApJ...679.1338R,2009Natur.457...63G} \footnote{Available at
\url{http://www.Dendrograms.org/en/latest/}} to quantify the clumpy structure of
the cloud. The \texttt{Dendrogram}s are representations of how the isosurfaces
in a 3-D PPV data cube nest inside one another. The
``Leaves'' of a \texttt{Dendrogram} correspond to the regions that have emission
enhancements in the 3-D PPV data cube, and they correspond to the ``clumps''
found by the well-known \texttt{clumpfind} algorithm
\citep{1994ApJ...428..693W}.
One advantage of using \texttt{Dendrogram} is that its results are less dependent on
technical parameters (for instance, the brightness temperature difference
between contours). In this work, the term {  cloud} is used to refer to the
whole cloud G052.24$+$00.74, and the term {  clump} is used to refer to the
sub-structures of the cloud identified by the \texttt{Dendrogram} algorithm.

\begin{figure}
\includegraphics[width=0.5 \textwidth]{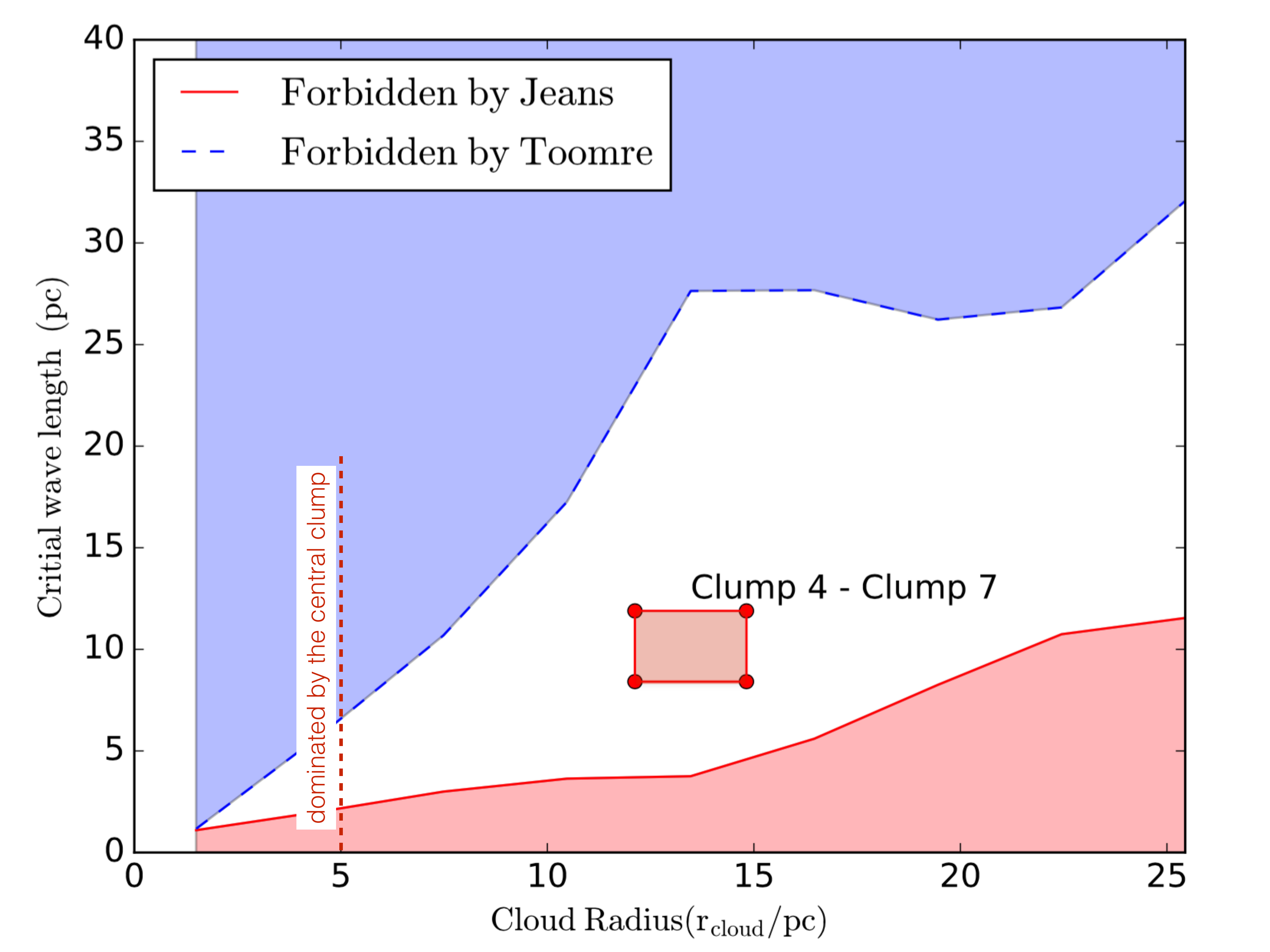}
\caption{Physical scales involved in the fragmentation process. 
The $x$-axis is the cloud radius, and the $y$-axis denotes the critical length
scales involved in the fragmentation process.
The blue dashed line stands for the Toomre length $l_{\rm Toomre} = 2 \pi G
\Sigma / \kappa^2$. In the blue shaded region fragmentation is
suppressed due to shear. The red solid line stands for the Jeans length $l_{\rm
Jeans} =  2 \sigma_{\rm v}^2 / G \Sigma$ where  we have chosen $\sigma_v = 
1\;\rm km/s$. In red shaded region fragmentation is suppressed due to thermal-turbulent
support. The clumps 4
and 7 form a pair at a cloud radius of 13.5 pc, and they are separated by 10
pc. It is indicated in this diagram as the red square, where the size of the
square represents the estimated uncertainty.
{   {  Since we did not de-project the clump separation into 3D, and the
cloud is inclined with an angle of $45^{\circ}$}, we estimate an uncertainty of
$\sqrt{2}$  (see text for details).
The central 5 pc is dominated by the central clump (measured from from Fig. \ref{fig:rings}),
where we do not expect our analysis to apply.
This is indicated by the vertical red dashed line.}
\label{fig:fragmentationi}}
\end{figure}

In this work, we smoothed the data in the velocity direction. The smoothed data
cube has a velocity resolution of 0.4 km/s and a rms noise level of $0.17$ K.
Then we identify clumps from the data using the \texttt{Dendrogram} program.
The program requires three inputs, the minimum value to consider in the dataset
(\texttt{min\_value}), and minimum difference for a structure to be considered
as independent (\texttt{min\_delta}), as well as the minimum number of pixels in
a given structure
 (\texttt{min\_npix}). We use \texttt{min\_value} =
0.67 K, \texttt{min\_delta} = 0.33 K and \texttt{min\_npix} = 16. This choice of
the parameters is relatively conservative, which ensures that only highly
significant structures are considered independent.  A change of these thresholds
to lower values produces a larger number of smaller clumps, but the significant
(e.g. these marked red in Fig. \ref{fig:fragmentationi}) structures remain
 unchanged. \footnote{  One can find in \citet{2014PhDT.......291L} results from
 a different combination of parameters. The clumps identified with these
 different parameter combinations are sometimes different, but the most
 significant clumps can always be robustly identified.}
 
 { 
 One should also note that the velocity dispersions of the extracted structures
 are dependent on the choice of the parameters. Structures identified with a
 relatively high threshold are more compact in 3D PPV space, and thus have
 smaller velocity dispersions.  This may be part of the reason why the
 total velocity dispersions (e.g.
 as in Fig. \ref{fig:mom}) estimated on lines of sights are larger than the
 velocity dispersion of the clumps.
}

\begin{figure}
\includegraphics[width=0.5 \textwidth]{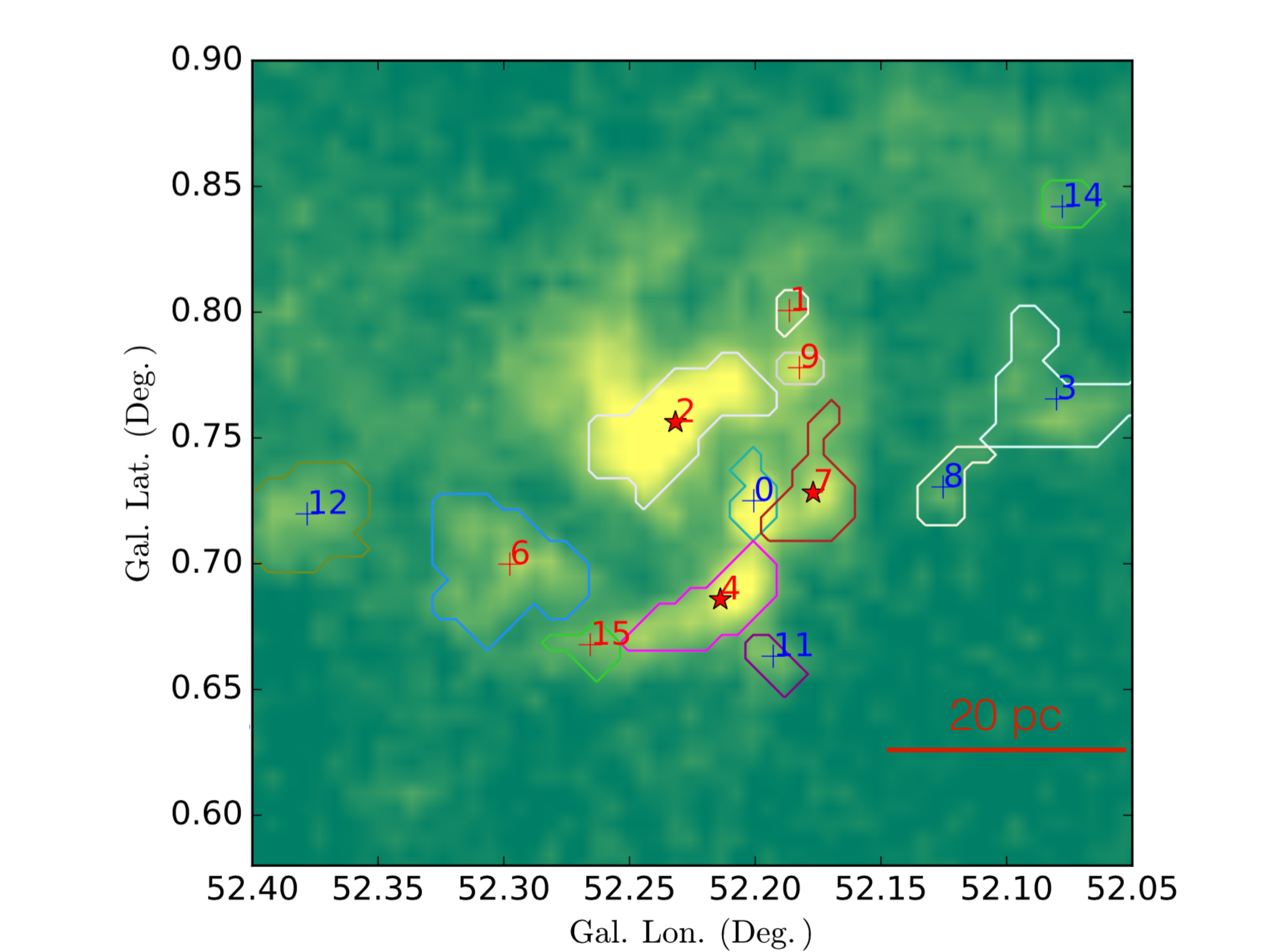}

\caption{Clumps (leaves) identified  by the \texttt{Dendrogram} method.
Here we use different colours to indicate different leaves for clarity.
Each leaf has a unique ID. We divide the leaves into three groups,
one group contains clumps that belong to the ``spiral arm'' of the
cloud (red numbers), and the other group contain clumps that are
outside the ``spiral arm'' (blue numbers). The third group contains clumps
2, 4 and 7, which exhibit clear evidences of ongoing star formation (red
stars).
\label{fig:leaves}}
\end{figure}

 In Figure \ref{fig:leaves} we plot the  {leaves} of
the \texttt{Dendrogram}. IDs of the leaves are also plotted. A detailed
catalogue can be found in Appendix \ref{sec:appendix:table}. 

\begin{figure}
\includegraphics[width=0.5 \textwidth]{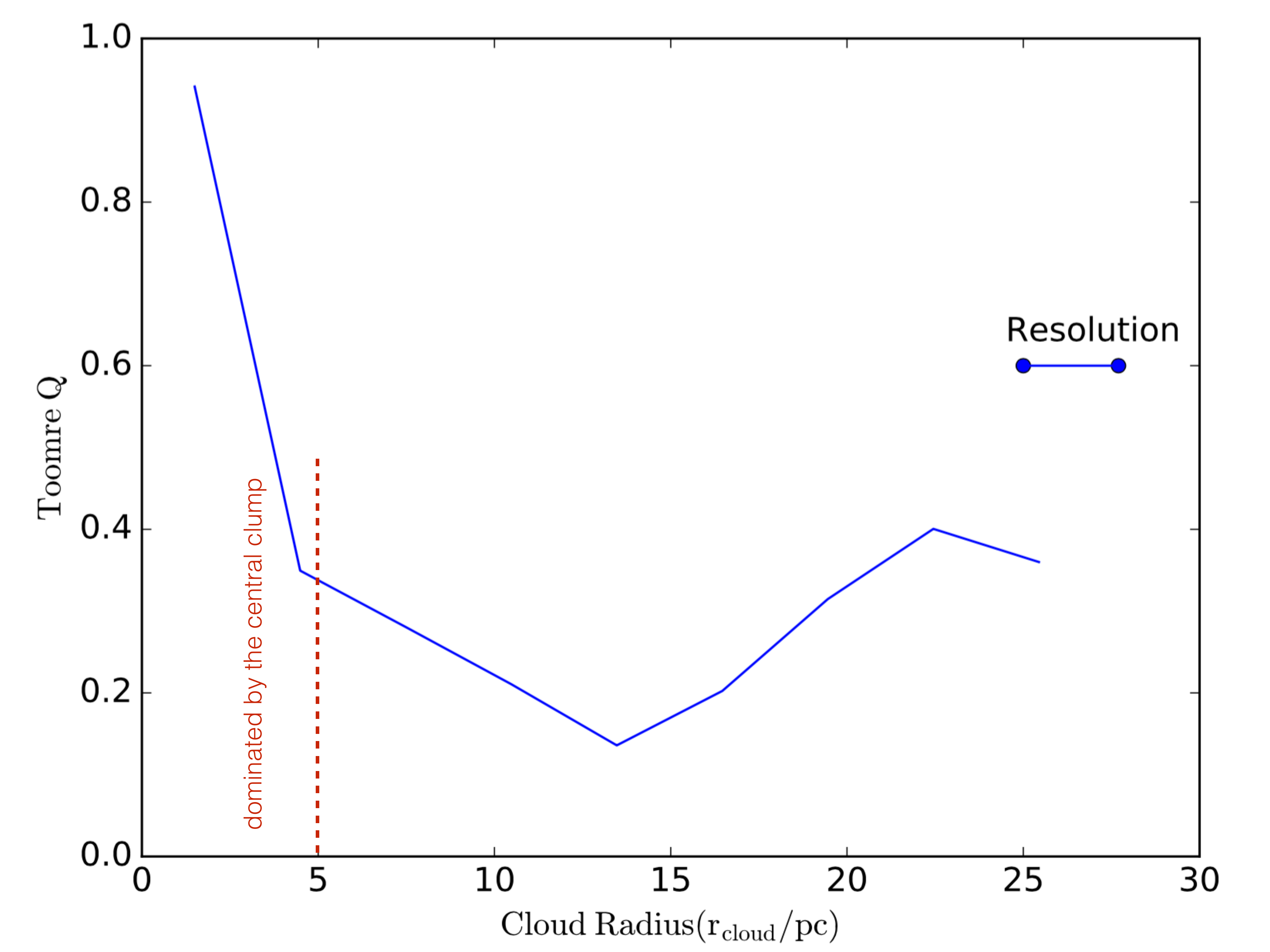}
\caption{Toomre Q parameter as a function of cloud radius ($r_{\rm cloud}$). The
Toomre Q is defined as $Q = \sigma_{\rm v} \kappa / \pi G \Sigma$ where 
$\sigma_{\rm v} = 1 \;\rm km/s$, $\Sigma$ is the ring-averaged column density
and $\kappa$ is the epicyclic frequency. The resolution of the plot is
 2.7 pc (limited by the size of the rings). It is indicated as the blue
 horizontal bar.
 The central 5 pc is dominated by the central clump.
This is indicated by the vertical red dashed line.}
\label{fig:toomre}
\end{figure}

\citet{1964ApJ...139.1217T} developed a theory for the stability of a disk of
stars. The formalism is general and has been applied to various systems that
are rotationally supported, such as disk galaxies and protostellar disks. In
this formalism, the stability of a disk is characterised by  the Toomre Q
parameter
\begin{equation}\label{eq:q}
Q = \frac{\sigma_{\rm v} \kappa}{ \pi G \Sigma}\;,
\end{equation}
{   where $\kappa$ is the epicyclic frequency,} $\sigma_{\rm v}$ represents
``internal'' supports such as thermal support and turbulence, and $\Sigma$ is
the surface density.
 If $Q > 1$ the disk is stable, and if $Q < 1$ the disk is unstable against
 perturbations and would fragment. We choose $\sigma_{\rm v} = 1\;\rm km/s$,
 which is the typical velocity dispersion to expect for clumps
 \citep{2015A&A...579A..91W} \footnote{  Which is consistent with the velocity
 dispersions of the clumps presented in Table \ref{table:clumps}.}.
In Fig.
\ref{fig:toomre} plot the Toomre Q as a function of disk radius. Except the
central region, the disk is unstable with $Q<1$. This is consistent with the
fact that the disk fragments into clumps due to gravitational instability.

In the formalism of \citet{1964ApJ...139.1217T} 
\citep[see also][]{2008gady.book.....B,1983MNRAS.203...31E}, the fragmentation
is determined by two length scales: the Toomre length and the Jeans length.
The Toomre length is defined as
\begin{equation}
l_{\rm Toomre} = \frac{2 \pi G \Sigma}{\kappa^2}\;.
\end{equation}
For disks with flat rotational profiles, $\kappa = \sqrt{2} \Omega$ where
$\Omega$ is the angular velocity, {  and $\Omega = v_{\rm circ } / r =
6.6\;\rm km/s / r_{\rm cloud}$ in our case.} This sets an upper limit to the
fragmentation length scale, and the growth of perturbations with $l > l_{\rm Toomre}$ are suppressed due to shear. The Jeans length  is defined as 
\begin{equation}
l_{\rm Jeans} = \frac{ 2 \sigma_{\rm v}^2}{ G \Sigma}\;,
\end{equation}
where $\sigma_v$ is the velocity dispersion, and it includes thermal and
non-thermal (e.g. turbulent) contributions. The Jeans length
is a lower limit to the fragmentation length, and growth of perturbations with
$l < l_{\rm Jeans}$ are suppressed due to thermal and non-thermal {  (e.g.
turbulent)} supports.  Only perturbations with $l_{\rm Jeans} < l <
l_{\rm Toomre}$ can be amplified.

The fragmentation length scale can be probed by studying the separations of the
neighbouring clumps.  In Fig. \ref{fig:fragmentationi} we plot these two
length scales as a function of cloud radius (defined in Sect. \ref{sec:structure}). 
% The the major uncertainty in our result comes form the uncertainty of the
% distance estimate. We adopt a typical uncertainty of $15 \%$
% \citep{2009ApJ...699.1153R}.
From the cloud, we
identify a pair of clumps  (4 -- 7 , see Fig.
\ref{fig:leaves}). Excluding the clump at the very centre, this is the
only pair of clumps that are significant on the ATLASGAL
\citep{2009A&A...504..415S} tile and {   thus covered in
subsequent studies \citep{2015A&A...579A..91W}.}
% And these are also indicated in Fig.
% \ref{fig:fragmentationi}. The clump pair 1--9 share a cloud radius of $\sim 10$
% parsec and are separated by $\sim 3$ parsec, and  
The clump pair 4--7 share a cloud
radius of 14 parsec and are separated by $\sim 10$ parsec. 
% From the location of
% the clump pairs in  Fig. \ref{fig:fragmentationi}. 
% The clump pair 1--9 are
% barely separated on the image, and this correspond to the fact that the
% fragmentation is suppresses by thermal-turbulent support as indicated in Fig.
% \ref{fig:fragmentationi}. 
They are well separated and both exhibit
signs of star formation. {  Since we did
not deproject the separation into 3D, we estimate an uncertainty of $\sqrt{2}$
of the clump separation estimate (because the cloud inclination is
$45^{\circ}$).
} In Fig. \ref{fig:fragmentationi}, this pair occupies a position where the growth of perturbations is allowed. The
predicted length scale of the fragmentation process using the formalism of
\citet{1964ApJ...139.1217T} matches well with the observed clump separation.
{  Interestingly, this pair also stays at a radius where the fragmentation is
most likely to occur (measured by the difference between $l_{\rm Jeans}$ and 
$l_{\rm Toomre}$). This supports our hypothesis that the Spiral Cloud fragments
due to gravitational instability.}

\subsection{Star formation in the clumps}
To further study the fragmentation process,
we divide the leaves
(clumps) into different groups.
First, we make a distinction between the clumps inside the spiral arm and
the clumps outside the spiral arm. Second, since three of the clumps (clump
2, 4 and 7) show evidences of star formation (inside clump 2, a star cluster
has already formed, and massive stars inside this star cluster are probably
triggering the formation of a next generation of stars. A bubble is found in
clump 7. RMS YSOs are found in clump 4 \citep{2012MNRAS.421..408T}), we
separate them from the rest of the clumps. Finally, we have three groups of
clumps. The first group include the clumps that exhibit obvious star formation
activities  (clump 2, 4 and 7), the second group include the other clumps
that are inside the spiral arm, and the third group consists of the rest of
the clumps.

We analyse the physical properties of the three different groups of clumps. In
deriving the properties, we use the same formalism as used in
\citet{2009Natur.457...63G}. The detailed properties of the clumps can be found
in Appendix \ref{sec:appendix:table}.
In Fig. \ref{fig:vir}, we plot $\sigma_{\rm v}^2/R$ versus $\Sigma$ for all the
clumps \citep[][where $\Sigma \equiv M/ \pi R^2$]{2001ApJ...551..747S}. Lines of
different virial parameters are derived from
\citep[][]{1992ApJ...395..140B}
\begin{equation}
\alpha_{\rm vir}=\frac{5 \sigma_{\rm v}^2 R}{G M}\;,
\end{equation}
where $\alpha_{\rm vir}$ is the viral parameter,
$\sigma_{\rm v}$ is the velocity dispersion of the clumps,
$R$ is the radii of the clumps, G is gravitational constant,
and $M$ is the mass clump.
The clumps that belong to the spiral arm are close to gravitationally bound
($0.5<\alpha_{\rm vir}<1$), while the other clumps are not.
Interestingly, all the three clumps that show star formation activities are
gravitationally bound.

\begin{figure}
\includegraphics[width=0.5 \textwidth]{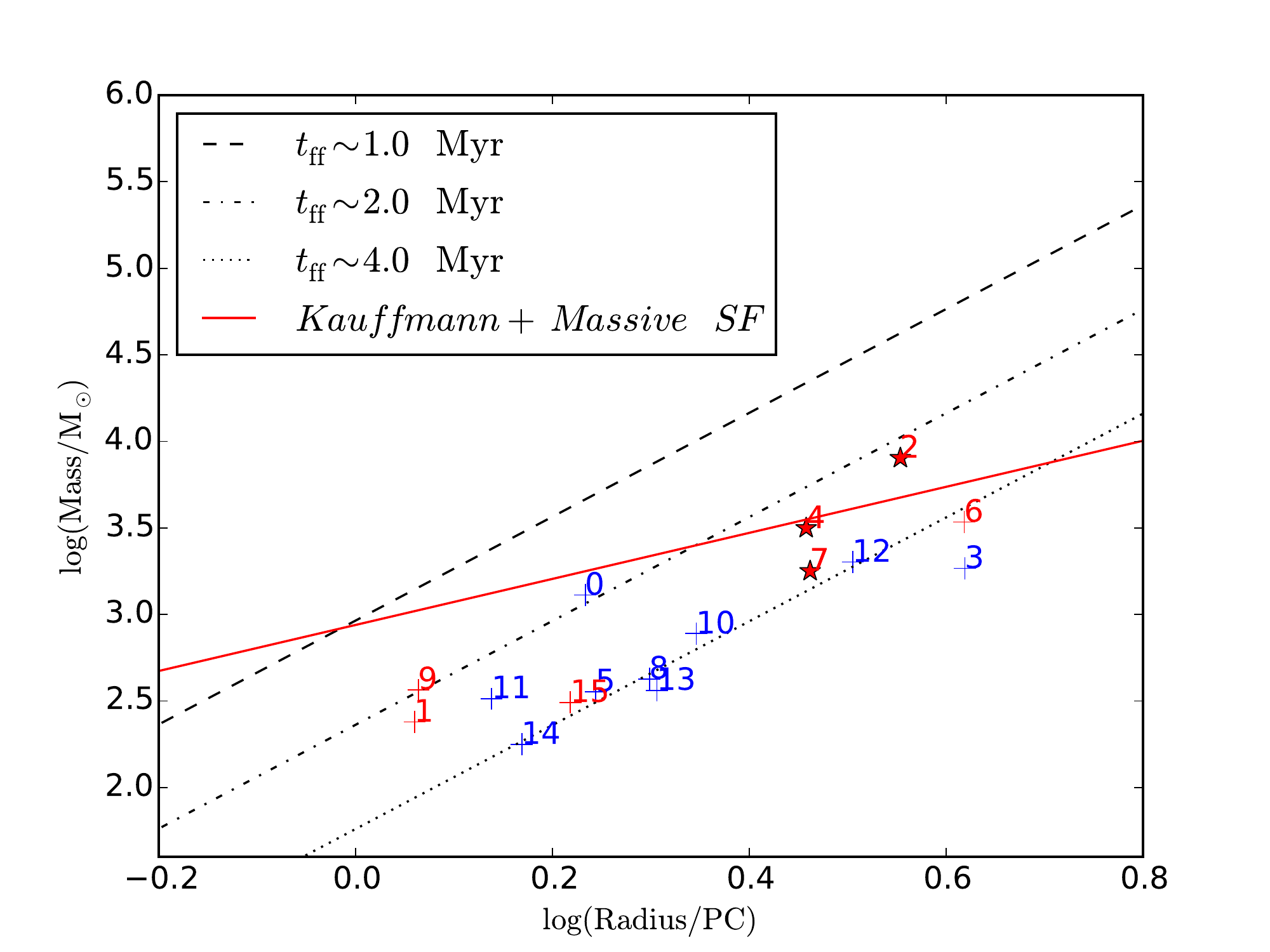}
\caption{Clump mass versus clump radius for all the clumps. 
We divide the clumps into three groups, one group contains the clumps that
belong to the spiral arm of the cloud (red crosses), and the other group contain the clouds that are outside the ``spiral arm''
(blue crosses). The third group contains clumps 2, 4 and 7, which exhibit
clear evidences of ongoing star formation (red stars with numbers). IDs of the
clumps are indicated. Lines of different free-fall time $t_{\rm
ff}\sim \rho^{-1/2}$ are included. The red solid line corresponds
to the threshold of massive star formation ${\rm M} = 870\,
{\rm M}_{\odot}  ({\rm r/pc)^{1.33}}$ derived in
\citet{2010ApJ...716..433K}. \label{fig:tff}}
\end{figure}

The correspondence of a small viral parameter and active star formation
 leads us to suggest that the star formation inside the clumps is
controlled largely by self-gravity. Another way to look at the problem is to
consider the free-fall timescale. This is a characteristic
timescale that governs the gravitational collapse. It is defined as $t_{\rm ff}\sim (G
\rho)^{-1/2}$, $\rho \equiv M/R^3$. In Figure \ref{fig:tff} we plot the masses
of the clumps against the radii of the clumps. Lines that correspond to
different free-fall timescales $t_{\rm ff}$ are added. The clumps that belong
to the spiral arm have significantly shorter free-fall timescales, and the three
clumps that are forming stars have not only short free-fall timescales but also
higher masses.

A star cluster has formed at the centre of clump 2 \citep{2005ApJ...635..560M}.
This clump is distinguished in two ways. First, it is the clump that resides
right at the centre of the spiral structure. Second, compared to all the other
clumps, it has the smallest viral parameter and the shortest free-fall
timescale. {  The connection between star formation and indicators of the
importance of self-gravity suggests that gravity is playing a determining role
in the evolution of the clumps.}
\citet{2010ApJ...716..433K} derived an empirical threshold ${\rm M} = 870\,
{\rm M}_{\odot}  ({\rm r/pc)^{1.33}}$ for the formation of
massive stars in the mass-size plane, {  e.g. clumps that are above this line are active in
massive star formation}. The results have been confirmed by studies of
larger samples,  e.g. ATLASGAL clumps \citep{2013MNRAS.435..400U}.
This threshold is also indicated in Fig.
\ref{fig:tff}.
Only clumps {  2, 4 and 7} host  \HII$\ $
 regions \footnote{Results from the Simbad
 \url{http://simbad.u-strasbg.fr/simbad/} database.}, and clumps 2 and 4 are above this threshold.
 {  However, we note that the result is true for these clump-like objects only in a statistical sense.
 }
% and
% this is consistent with the existence the HII regions in these clumps.

\begin{figure}
\includegraphics[width=0.5 \textwidth]{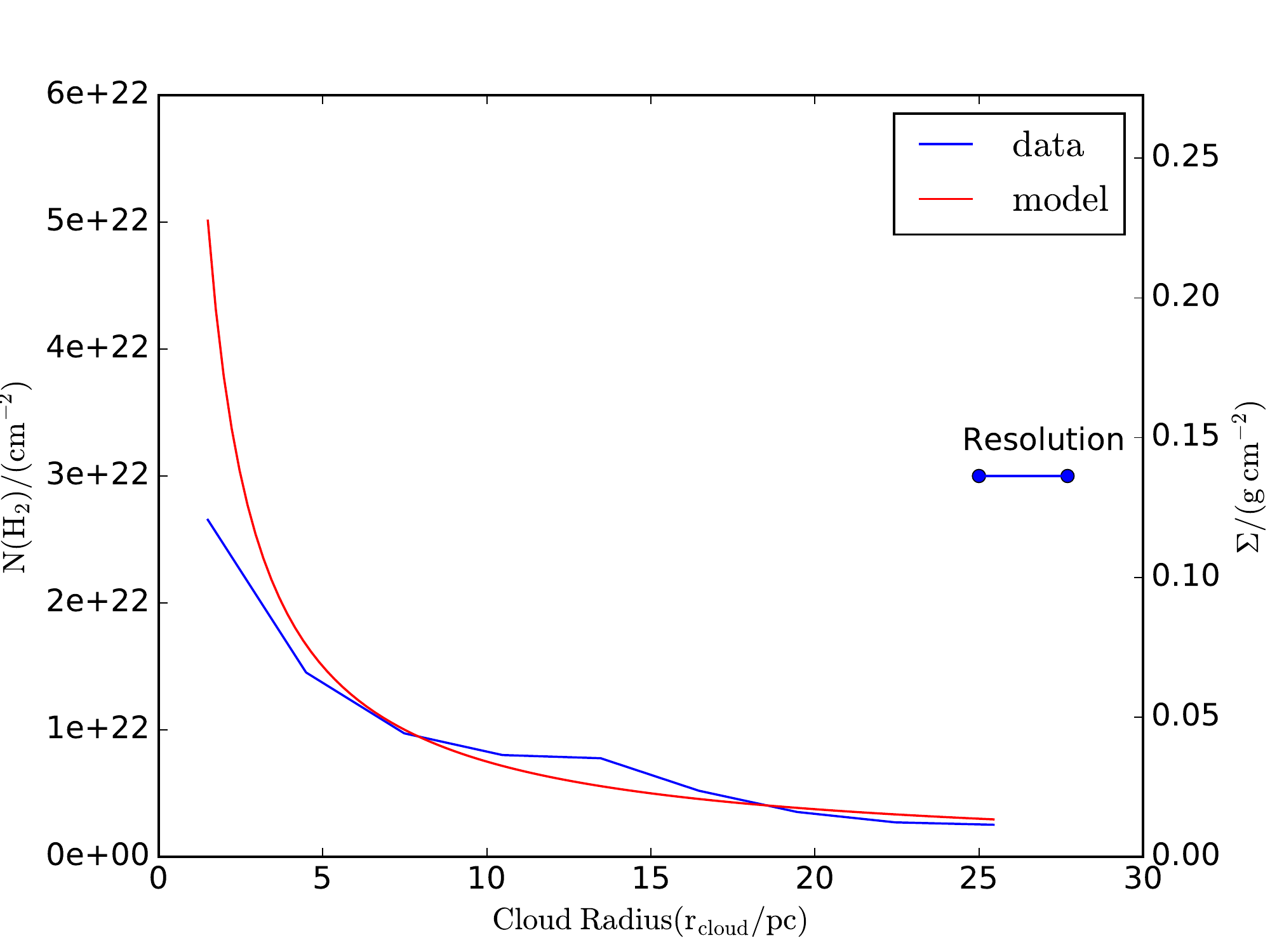}
\caption{ Column density structure of the Spiral Cloud. The figure presents 
ring-averaged column density as a function of cloud radius. The blue line stands
for the data and the red line stands for the analytical model (Eq.
\ref{eq:sigma:r}).
% { 
% Right Panel:} Velocity structure of the cloud. The green line stands for the
% velocity dispersion of gas inside different rings, and the blue line stands for
% the theoretically-expected velocity dispersion assuming virial balance. The two
% red dashed lines stand for the theoretically-derived rotation velocity
% assuming the analytical model.  They correspond to a perfect edge-on
% disk and a disk inclined at 45$^{\circ}$, respectively.
{  The telescope resolution is $\sim
2.2\;\rm pc$ at a distance of $9.8\;\rm kpc$. The resolution of the plot (2.7
pc, limited by the widths of the rings) is indicated as the blue horizontal
bar.} See Sect.
\ref{sec:structure} for details.
\label{fig:cloud:structure}}
\end{figure}

\begin{figure}
\includegraphics[width=0.5 \textwidth]{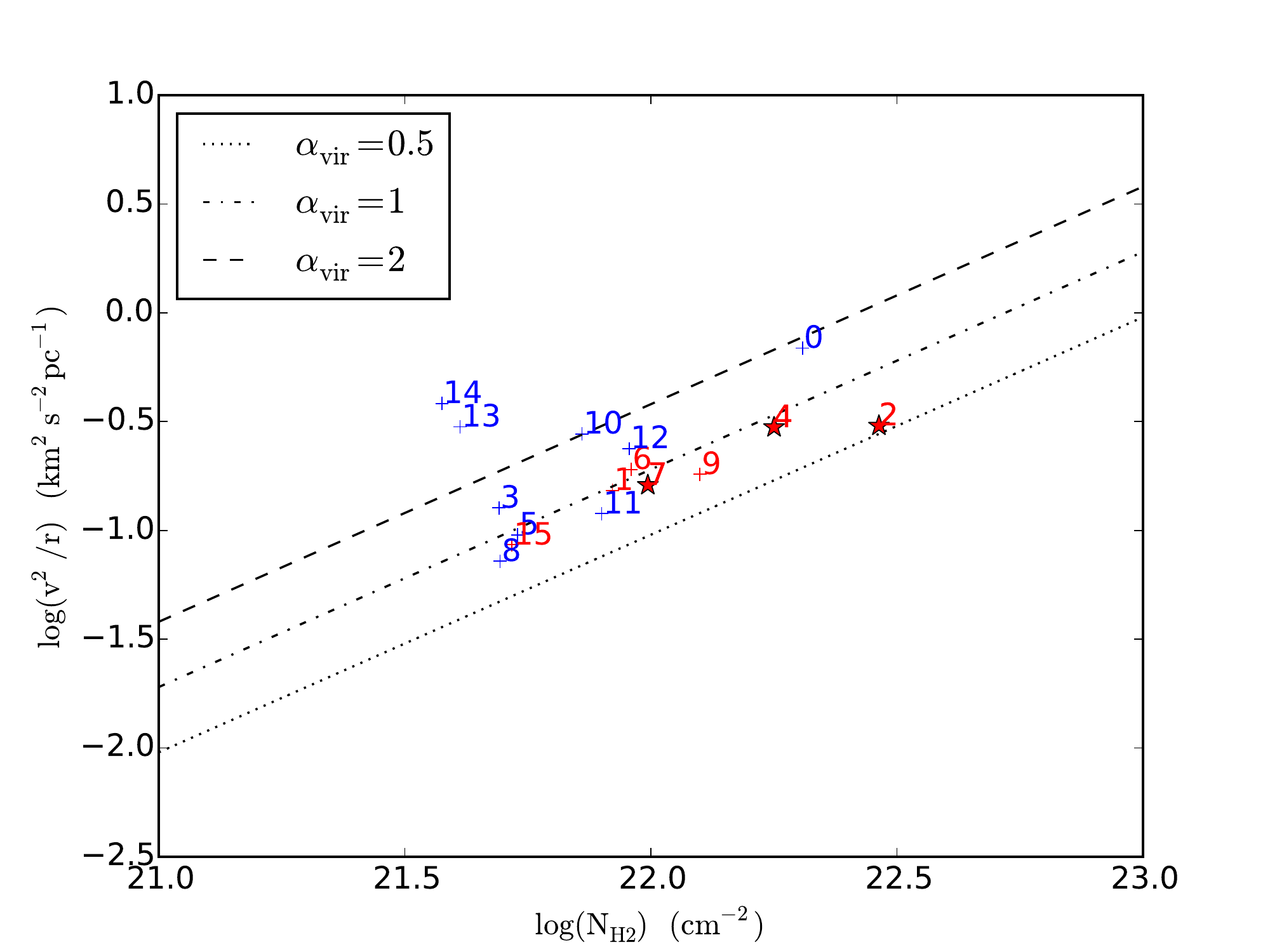}
\caption{Ratio of $\sigma_v^2/R$ as a function of clump column density $N_{\rm H_{\rm 2}}$. 
We divide the leaves into three groups, one group contains the clumps that belong to the spiral arm
of the cloud (red crosses), and the other group contains clumps that are
outside the spiral arm (blue crosses).
The third group contains clump 2,4 and 7, which exhibit clear evidences of
ongoing star formation (red stars with numbers). IDs of the
clumps are indicated
Lines of different virial parameters $\alpha_{\rm vir}=5 \sigma_{\rm v}^2 R/G \Sigma $ are included.\label{fig:vir}}
\end{figure}

\section{Rotating molecular clouds in observations and
simulations}
How do molecular clouds evolve? It is widely recognised that cloud evolution
is governed by the combined effects of (ordered or disordered) kinematic motion,
turbulence, gravity and magnetic field. However, the relative importance of
these factors is not known. The importance of rotation in cloud
evolution has long been recognised \citep{1999A&AS..134..241P} but remains
poorly-constrained.

In the case of the Spiral Cloud, rotation seems to play an important role.
The dynamics of the cloud can be reasonably described as a balance between
rotation and gravity, the interplay of which regulates the fragmentation
process.

 Is rotation playing a role in molecular clouds in
general? The idea of rotating clouds is not new
\citep{1983A&A...119..109B,1984MNRAS.210...43Z,
1978ApJ...224..488B,1999A&AS..134..241P}.
However, from the current literature it is still unclear if rotation is playing a role in the Milky Way molecular clouds. 
% The lack of observational constraints can be
% attributed to the fact that observations of the Milky Way molecular cloud
% suffer from line of sight confusions, which makes it difficult to study rotation
% on a statistical basis. An easier way to study cloud rotation is to study the
% molecular clouds in nearby Galaxies.
 Interestingly, the possibility of rotating
molecular clouds has been suggested by several groups from observations of
different galaxies \citep{2007ApJ...654..240R,2011ApJ...732...79I}.
Recently, \citet{2015ApJ...803...16U} reported  CARMA observations of the
lenticular galaxy NGC4526 in $^{12}$CO line where the authors found the
angular momentum vectors of the molecular clouds tend to align with the minor
axis of the galaxy. { This leads the authors to conclude that the clouds are
rotating.} Our Spiral Cloud
could be representative of a population of rotating molecular clouds in our
galaxy.

% Galactic-scale simulations as well as analytical works also provide hints
% concerning cloud rotation. In the simulations of galaxy disks, frequent
% encounters between molecular clouds are expected and observed
% \citep{2000ApJ...536..173T,2002ApJ...570..132K,2006MNRAS.367..873D,2006MNRAS.371.1663D,2006ApJ...647..997S,2009ApJ...700..358T, 2012MNRAS.420.3490C}.
% The encounters between molecular clouds produces spiral-shaped
% rotating clouds and elongated gas filaments,  as well as complexes that
% contain both. Our Spiral Cloud is connected with a neighbouring cloud with
% some wispy gas  filaments (Figure \ref{fig:channel5152}). This {  spiral
% cloud-filament} seems to resemble the structures seen in these numerical
% simulations.

% Our Spiral Cloud is connected with a neighbouring cloud with some wispy gas
% filaments (Figure \ref{fig:channel5152}). This {  spiral cloud-filament}
% configuration has also been in many simulations
% \citep[e.g.][]{2011ApJ...730...11T,2011MNRAS.413.2935D,2012arXiv1211.1681V,2014MNRAS.439..936F,2012MNRAS.420.3490C,2015MNRAS.448.1007B}.

Rotation can be important also at clump and core scales
\citep{1993ApJ...406..528G}. In our case, it is possible that a  significant
amount of angular momentum of the cloud will be transferred to the clumps
\citep{1984MNRAS.210...43Z}. The $^{13}$CO emission is optically thick
at these scales and we are not able to test this hypothesis. 
However, recent results do indicate that rotation can become dominant at $\sim$
pc scale \citep{2015ApJ...804...37L}.
% Gravity plays an important role in controlling the subsequent star formation
% processes that occur in the clumps. The clumps that forms stars are massive
% ($M\geq100 M_{\odot} $), gravitationally bound ( with viral parameter
% $\alpha_{\rm vir}\leq 1$), and have short dynamical time-scale $t_{\rm dyn} \leq
% 1\;\rm Myr$. The clump 22 that harbours the star cluster is the most massive
% ($\sim 5000 M_{\odot}$) clump with the smallest viral parameter ($\alpha_{\rm
% vir}\sim 0.5$).

Even though turbulence does not seem to be dominant in shaping the structure of
the cloud on the large scale, it may play important roles in controlling the
fragmentation of the individual clumps. Indeed, all the clumps that belong to the spiral arm
exhibit supersonic line widths. It is possible that these clumps are supported
by turbulent motion, and turbulence in the clumps can be generated
during the fragmentation process \citep{2010A&A...520A..17K}. Further observations with
higher angular resolutions are needed, in order to disentangle  the roles of
turbulence and rotation on the subsequent fragmentation of the clumps.

% The clump 2 stays at the very center of the spiral cloud and shows prominent
% star formation. What triggers the star formation? While gravitational
% fragmentation can offer an explanation, it is also possible that the the
% star formation is triggered by cloud-cloud interaction
% \citep{2015arXiv150905287B,2015ApJ...806....7T,2011A&A...528A..50D,2014ApJ...780...36F,2009ApJ...696L.115F,2009ApJ...700..358T}.
% Indeed, in numerical simulations, encounters between different clouds 
% often result in spiral-shaped structures. It is possible that the cloud acquires
%  angular momentum during the merger of two smaller clouds. 

\section{Conclusions}
We report the study of a spiral-shaped molecular cloud in our Galaxy.

{  The cloud belongs to a large 500 pc gas filament
\citep{2013A&A...559A..34L}, which is stretching over $\sim 100 \;\rm pc$ above
the Galactic disk midplane.  }
The cloud exhibits a spiral-shaped morphology. It shows a velocity shift of
 $\sim 10\;\rm km/s$ at a scale of $\sim 30\;\rm pc$.  {  The observed
 kinematic structure can be reproduced if the cloud is
 rotationally supported.}
This lead us to conclude that on the cloud scale, rotation is important in
balancing against gravity.
The cloud is rotating {  in prograde direction with respect to the bulk of the
Milky Way.}

We analysed the dynamics and the fragmentation process under the framework of
gravitational instability of \citet{1964ApJ...139.1217T}. We found that our
cloud is unstable against gravitational collapse.
By analysing the cloud structure  in detail we found
that the separation between the clumps can be consistently reproduced assuming gravitational
instability. 

We studied the physical properties of the fragments, and found that the clumps on
the spiral arm part of the cloud are close to gravitationally bound. Star
formation occurs in the clumps that are gravitationally bound with short
free-fall times.
All the facts seem to indicate that gravitational instability is crucial for
the fragmentation of the cloud and self-gravity is driving the subsequent star
formation. 
When viewed against observations of cloud rotation in other Galaxies, we
speculate that our cloud could represent a category of
rotationally-supported clouds for which the interplay between gravity and
rotation plays a determining role in their evolution. 

\begin{acknowledgements} 
Guang-Xing Li received support for this research through a stipend
from the International Max Planck Research School (IMPRS) for Astronomy
and Astrophysics at the Universities of Bonn and Cologne and is currently
supported by the Priority Program 1573 ISM-SPP from the DFG (Deutsche
Forschungsgemeinschaft).
Guang-Xing Li would like
to thank ISIMA (International Summer Institute for Modelling in Astrophysics,
now Kavli Summer Program in Astrophysics) for facilitating collaborations, and
would like to thank Nicolas Peretto and Patrick Hennebelle for discussions.
Lee Hartman is acknowledged for giving a nice lecture in ISIMA, which motivates the thinking of gravity,
and Di Li is acknowledged for providing references on angular momentum.
Guang-Xing Li thanks Alexei Kritsuk for discussions and for his final push
towards the publication of the results. Guang-Xing Li thanks Manuel Behrendt for
discussions of gravitational instabilities and clumps in disk galaxies.
 
This publication makes use of molecular line data from the Boston University-FCRAO
Galactic Ring Survey (GRS). The GRS is a joint project of Boston University and
Five College Radio Astronomy Observatory, funded by the National Science
Foundation under grants AST-9800334, AST-0098562, \& AST-0100793.
This work is based in part on observations made with the Spitzer Space Telescope,
which is operated by the Jet Propulsion Laboratory, California Institute of Technology
under a contract with NASA. We also would like to thank Thomas Robitaille for
making his code publicly available. The research made use of the
\texttt{astropy} package. 

The referee has to be acknowledged, whose comments have lead to very significant
improvements in our paper.

\end{acknowledgements}

\begin{appendix}
%\subsection{Clump mass and Velocity}
%\section{Distance determination}

\section{Channel Maps of the Molecular Complex at $l\sim 52$}\label{appendix:B}
\begin{figure*}
\includegraphics[width=0.95 \textwidth]{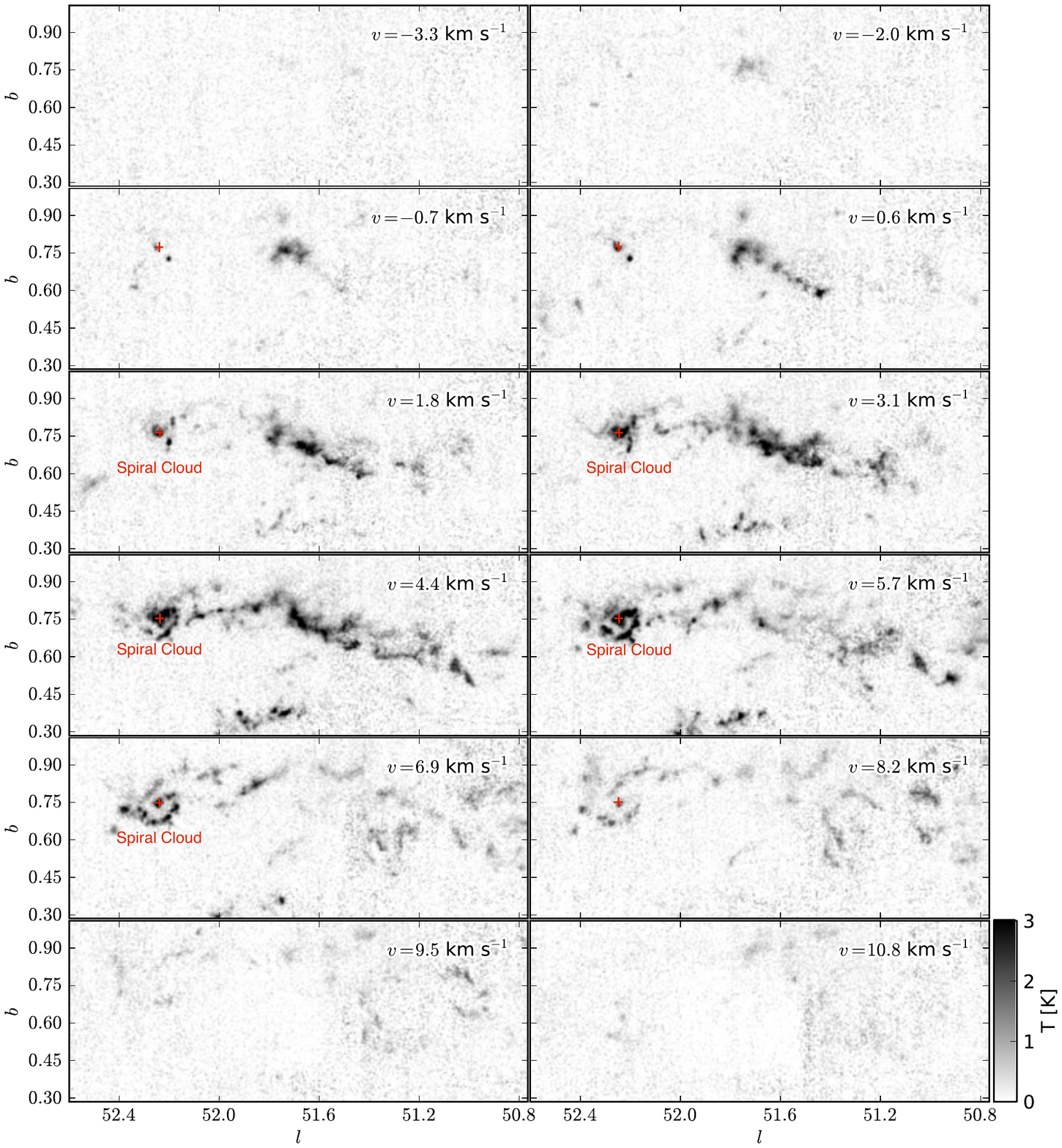}
\caption{$^{13}$CO(1-0) channel maps of the cloud complex. {  The red ``$+$''
marker indicate the centre of the Spiral Cloud.} The cloud complex belongs
to a 500-pc filamentary gas wisp, which has been reported in \citet{2013A&A...559A..34L}.
\label{fig:channel5152}}
\end{figure*}

In Figure \ref{fig:channel5152} we present the $^{13}$CO(1-0) channel maps of
the molecular cloud complex from the GRS survey \citep{2006ApJS..163..145J}.

\section{Physical Properties of the Clumps}\label{sec:appendix:table}\
The physical properties of the clumps are evaluated using the formalism outlined
in \citet{2009Natur.457...63G}. 
% The luminous mass is calculated according to
% $M_{\rm 13CO}=X_{\rm 13CO}L_{\rm 13CO}$ where $X_{\rm 13CO}=4.92\times
% 10^{20}\rm\; K\; km\;s^{-1}$ \citep{2001ApJ...551..747S} assuming optically thin
% emission, an excitation temperature of 10 K, and an X($^{13}$CO/$\rm H_2$) of
% $1.7\times 10^{-6}$.
The radii of the clumps are evaluated as $R=1.91 \times \sqrt{\sigma_{\rm
maj}\sigma_{\rm min}}$ where $\sigma_{\rm maj}$ and $\sigma_{\rm min}$ are the
major and minor axis dispersion \citep{1987ApJ...319..730S}. {  The velocity
dispersions of the clumps are evaluated using the intensity-weighted
standard deviation of velocities of all vorxels that belong to the clumps.} The column
densities of the clumps are evaluated as $\Sigma=M/\pi R^2$.
The physical properties of the clumps are listed in Table \ref{table:clumps}. 
% {  We found three matches with
% \citet{2015A&A...579A..91W}: Our clump 2 matches with G52.24$+$0.75, and the
% mass estimated from ATLASGAL is $6\times 10^3M_{\odot}$. }

\begin{table*}
\caption{Properties of the $^{13}$CO Clumps}              % title of Table
\label{table:1}      % is used to refer this table in the text
\centering                                      % used for centering table
\begin{tabular}{c c c c c c c c c}          % centered columns (4 columns)
\hline\hline                        % inserts double horizontal lines
ID & Mass & Gal. Lon. & Gal. Lat. &\textit{v}$_{\rm lsr}$ & Radius&  Velocity
Dispersion &Viral Parameter & Spiral arm  \\
 & ($M_{\odot}$)& (Degree) & (Degree)& ($\rm
 km\;s^{-1}$) &(parsec) &  ($\rm
 km\;s^{-1}$)  &\\   % table heading
\hline                                   % inserts single horizontal line
0&\num[round-precision=2,round-mode=figures,scientific-notation=true]{1.3e+03} & 52.20&0.73&0.94&1.76&1.10&1.8&False\\
1&\num[round-precision=2,round-mode=figures,scientific-notation=true]{2.4e+02} & 52.19&0.80&2.37&1.18&0.42&1.0&True\\
2&\num[round-precision=2,round-mode=figures,scientific-notation=true]{8.0e+03} & 52.23&0.76&4.33&3.67&1.05&0.6&True\\
3&\num[round-precision=2,round-mode=figures,scientific-notation=true]{1.8e+03} & 52.08&0.77&3.75&4.27&0.74&1.4&False\\
4&\num[round-precision=2,round-mode=figures,scientific-notation=true]{3.2e+03} & 52.21&0.69&5.02&2.95&0.94&0.9&True\\
5&\num[round-precision=2,round-mode=figures,scientific-notation=true]{3.6e+02} & 52.00&0.78&4.03&1.80&0.41&1.0&False\\
6&\num[round-precision=2,round-mode=figures,scientific-notation=true]{3.4e+03} & 52.30&0.70&5.81&4.26&0.90&1.1&True\\
7&\num[round-precision=2,round-mode=figures,scientific-notation=true]{1.8e+03} & 52.18&0.73&5.52&2.97&0.69&0.9&True\\
8&\num[round-precision=2,round-mode=figures,scientific-notation=true]{4.2e+02} & 52.13&0.73&4.76&2.04&0.38&0.8&False\\
9&\num[round-precision=2,round-mode=figures,scientific-notation=true]{3.7e+02} & 52.18&0.78&5.08&1.19&0.46&0.8&True\\
10&\num[round-precision=2,round-mode=figures,scientific-notation=true]{7.8e+02} & 52.01&0.86&5.54&2.28&0.79&2.1&False\\
11&\num[round-precision=2,round-mode=figures,scientific-notation=true]{3.3e+02} & 52.19&0.66&5.39&1.41&0.41&0.8&False\\
12&\num[round-precision=2,round-mode=figures,scientific-notation=true]{2.0e+03} & 52.38&0.72&6.44&3.28&0.88&1.4&False\\
13&\num[round-precision=2,round-mode=figures,scientific-notation=true]{3.6e+02} & 52.42&0.63&6.76&2.08&0.79&4.0&False\\
14&\num[round-precision=2,round-mode=figures,scientific-notation=true]{1.8e+02} & 52.08&0.84&6.44&1.51&0.76&5.5&False\\
15&\num[round-precision=2,round-mode=figures,scientific-notation=true]{3.1e+02} & 52.27&0.67&6.96&1.70&0.38&0.9&True\\
\hline                                   % inserts single horizontal line

\end{tabular}\label{table:clumps}

\end{table*}
% \section{Evaluation of Velocity Gradient\label{sec:ap:vg}}
% Based on the clumps (``leaves'') found out by the {  dendrofind} algorithm from the $^{13}$CO data cube, we evaluate the velocity gradient of each clump. 
% 
% 
% One ``leaf'' found out by the {  dendrofind} algorithm consists of  a list of positions in the 3D PPV data cube, each position $(x_i, y_j, v_k)$ have an intensity $I(x_i, y_j, v_k)$. At one spatial position $(x_i, y_j)$, there are several velocity channels that belong to one ``leaf''. At one spatial position $(x_i, y_j)$, we evaluate the intensity-weighted velocity  using
% \begin{equation}
% v_{ij}\equiv v(x_i,y_j)=\frac{\sum_k I(x_i, y_j, v_k) \times v_k}{\sum_k I(x_i, y_j, v_k) }\;.
% \end{equation}
% A linear regression is applied on each point set $v_{ij}$ to evaluate the velocity gradient.

\section{Analysis of the structure}\label{sec:appen:vg}
\begin{figure}
\includegraphics[width=0.5 \textwidth]{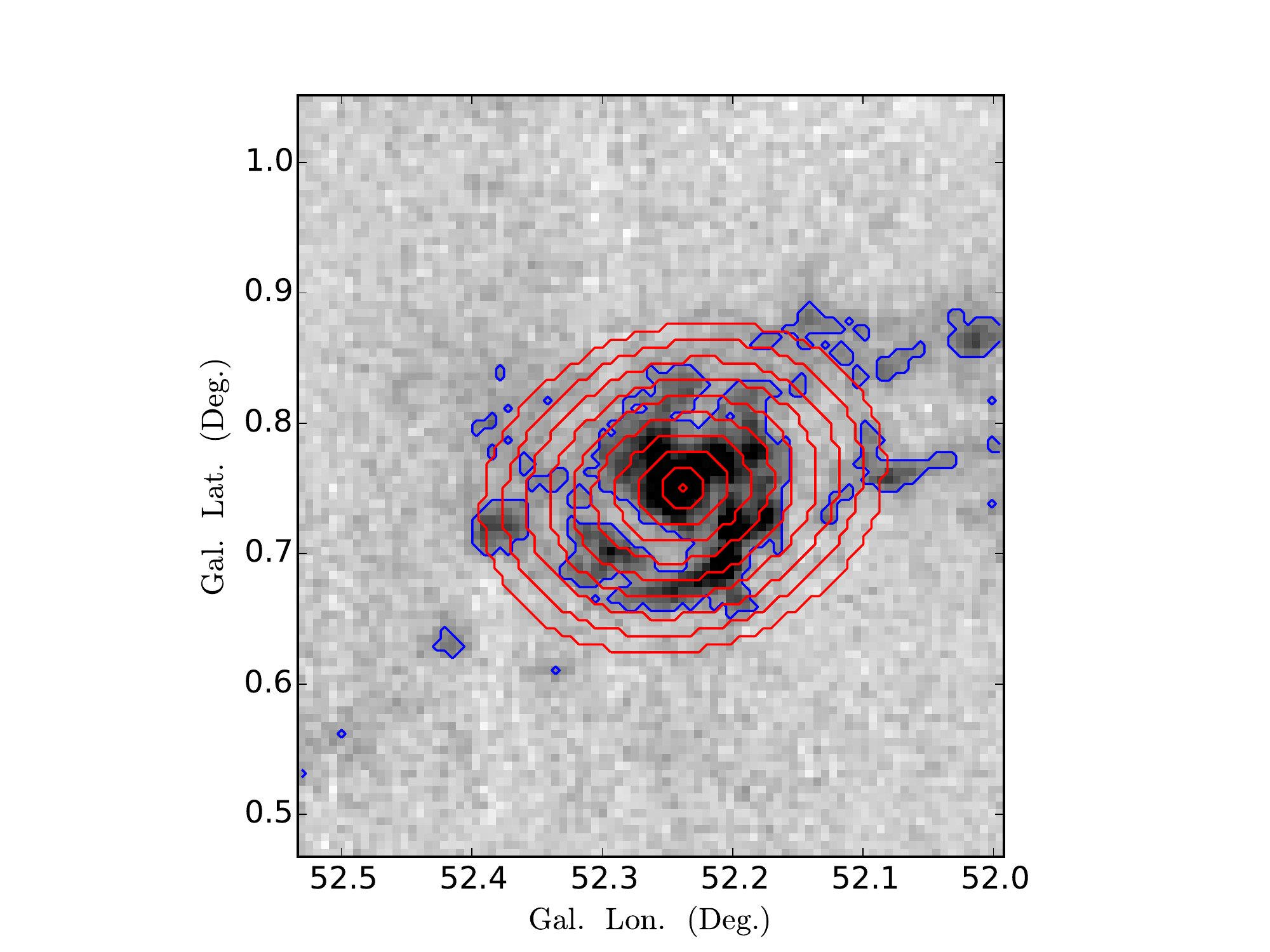}
\caption{A demonstration of our ring-based analysis of the cloud. {  The
spacing of the rings are 2.7 pc, which is comparable to the beam size (which is 2.2
pc)}. The blue solid lines stand for the region within which the emission is
analysed. The cut was made at $\int T_{A*} {\rm d}{\rm v} =  5.1 \;\rm km/s$,
{  This corresponds to 3.5 times the rms noise level.} The red contours denote
the boundaries of the rings that we use to divide and analyse the kinematic structure of the Spiral Cloud.
\label{fig:rings}}
\end{figure}

\begin{figure*}
\includegraphics[width=1 \textwidth]{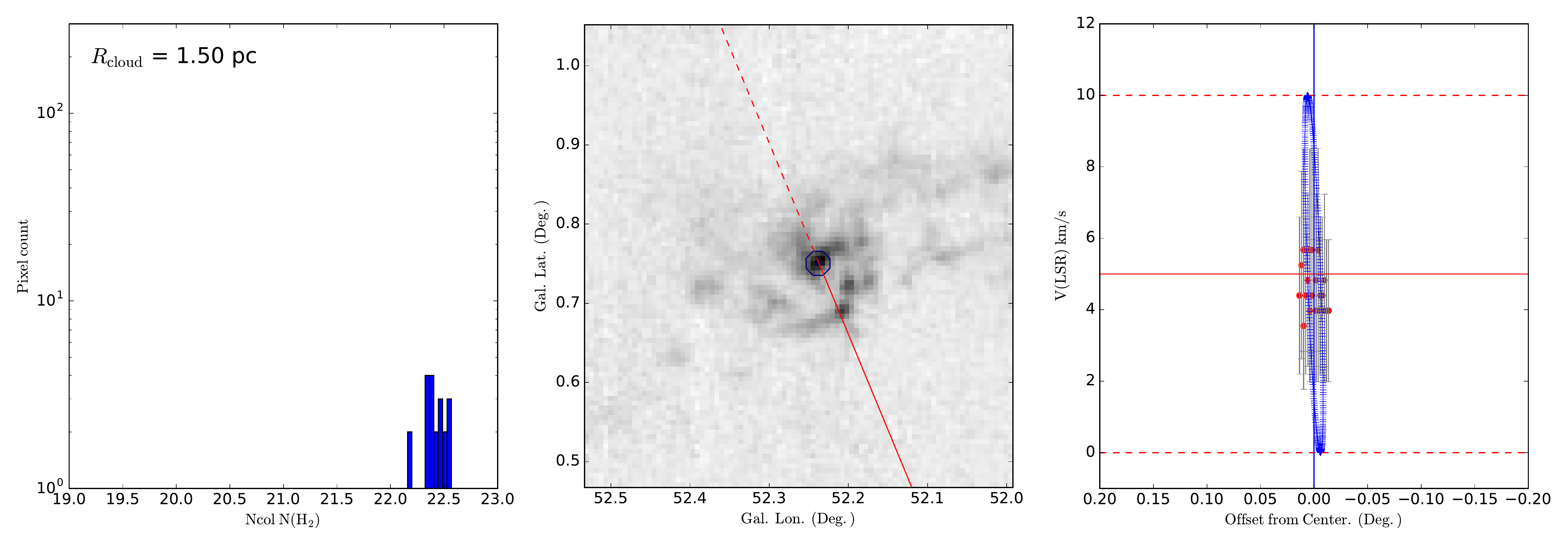}
\includegraphics[width=1 \textwidth]{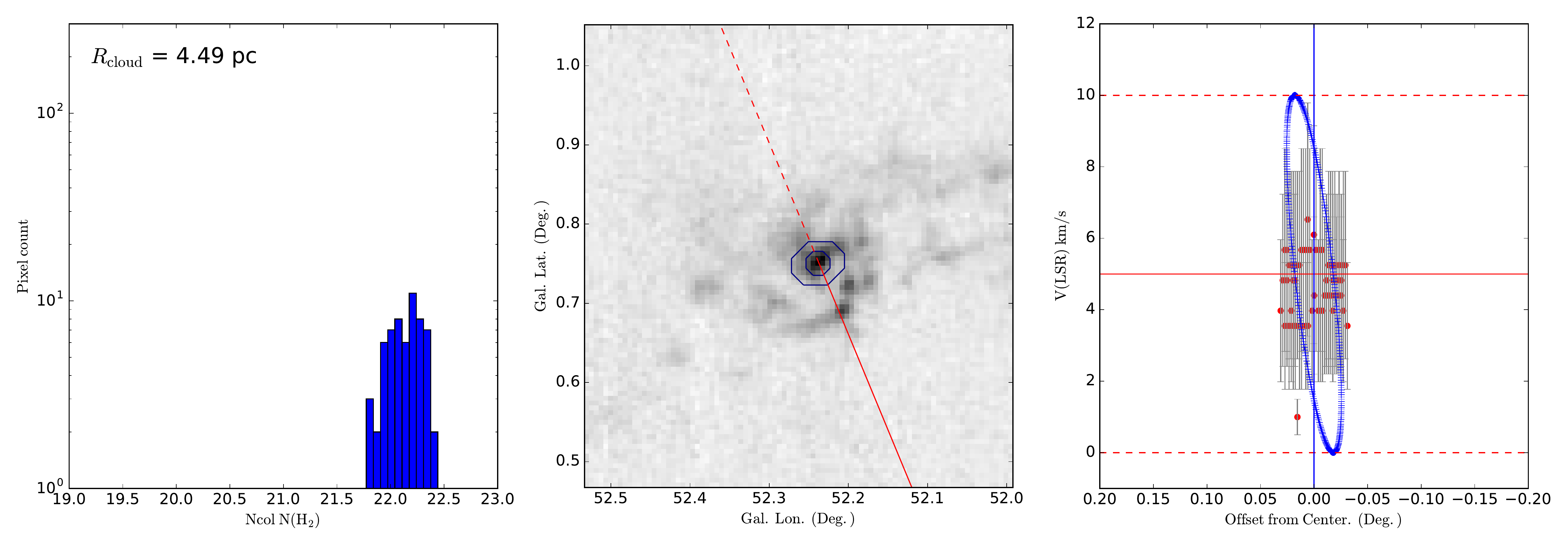}
\includegraphics[width=1 \textwidth]{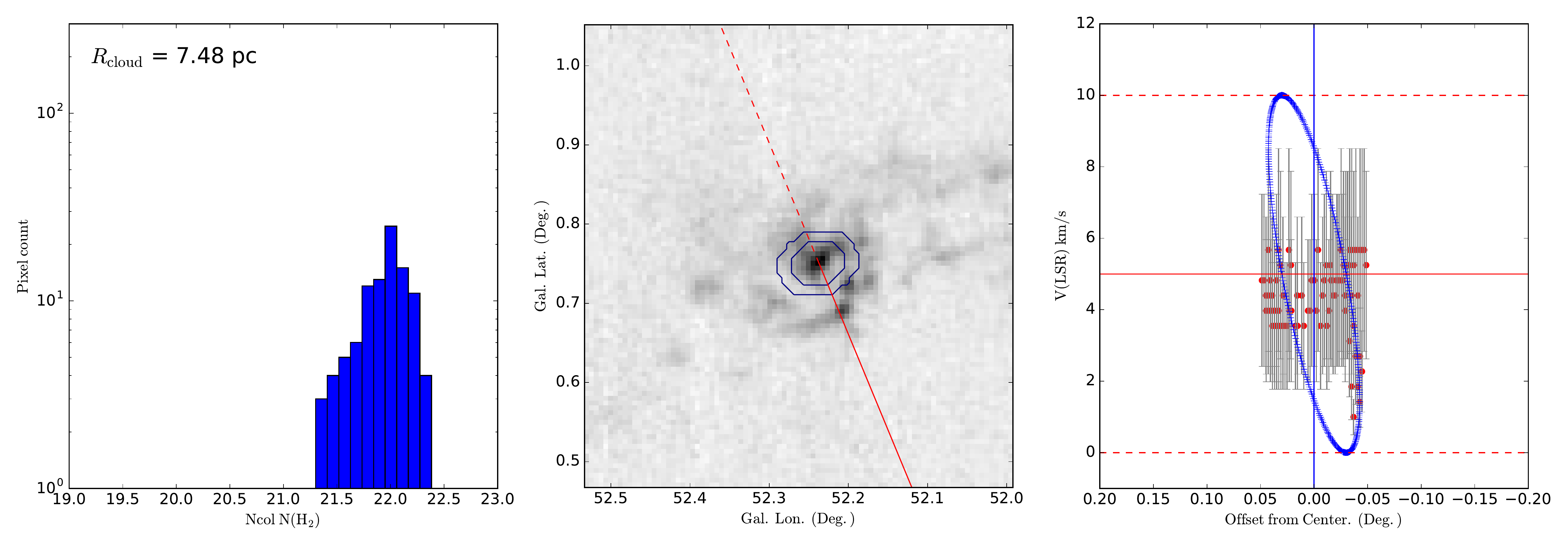}
\caption{Structure of the cloud viewed within rings.
On the left panels, we present the
distributions of estimated $\rm H_2$ column densities of the pixels.
 The mean radii of the rings ($r_{\rm cloud}$) are indicated  at the
 upper left corners of each panel.  {  In the middle panels, we plot the
boundaries of the rings.} {  The proposed
rotational axis is represented by the red lines. The red solid lines represent
parts of the rotational axes that are in front of the cloud, and the red dashed
lines represent parts of the rotational axes that are at the back of the cloud.
} the right panels we present the velocity centroids of the emission in the position-velocity space. {  Here, we have projected individual line of sight observations onto the major axis of ellipsoid
representing the cloud, and the x-axes are the projected offests from the
centres.} The horizontal lines in the right panels denote the range of
velocities within which the gas is gravitationally bound to the cloud. For each line of sight, we plot the velocity centroids and the FWHMs. {  The centroids of the emission
lines are represented as the red symbols, and the FWHMs of the of the emission
lines are represented as the errorbars. The horizontal red lines  represent the
systemic velocity of the cloud, which is $5\;\rm km/s$. The blue curve is the expected rotational
signature assuming the balance between rotation and gravity
(Sec.
\ref{sec:structure})}. The vertical blue lines in the middle and right panels indicate the centre of the cloud.
This figure is to be continued in the next page.
\label{fig:rings1}}
\end{figure*}

To analyse its structure we divide the cloud into different rings. The
positions of the rings are presented in Fig. \ref{fig:rings}, and in Fig.
\ref{fig:rings1} we present the velocity structure of the cloud inside
these rings.

We adopt a systemic velocity of 5 $\rm km/s$ for the Spiral Cloud.
For each position, the FHWMs of the emission lines are determined by ${\rm FWHM}
\approx  0.93 I / T_{\rm peak}$ where $I$ is the integrated intensity, and
$T_{\rm peak}$ is the peak the emission line. 

We also compare the observational data with a model of a flat rotating disk
model \citep{1963MNRAS.126..553M}, described in detail in Sect.
\ref{sec:structure}.
The model has a constant circular
velocity (flat rotation profile), characterised by $v_{\rm circ}$, which is 6.6
$\rm km/s$ (Eq. \ref{eq:v:virc}). Assuming an inclination of 45$^{\circ}$, we
derive the observed velocity signature of such a model.

{  At the very inner regions (up to $r_{\rm cloud} = 7.48\;\rm pc$), the
observed lines are sometimes not purely Gaussian. This might lead to some underestimates
of the column density, and might make the velocity centroid estimates
inaccurate.
Indeed, the line profiles are more irregular  at the region
$l\approx52.26^{\circ}$, $b\approx0.76^{\circ}$ where the line widths are much
higher (Fig. \ref{fig:mom}). A few causes for the irregular line profiles are
possible, such as line of sight superposition and opacity broadening. If the
latter is dominating, the deviation of our model from the model of
\citet{1963MNRAS.126..553M} would be smaller than Fig. \ref{fig:cloud:structure}
suggests.

Another uncertainty in our modelling is the distance. The expected rotational
velocity is linked to $r_0$ by Eq. \ref{eq:v:virc}, and $r_0$ is
proportional to the distance. Since the distance used in our analysis is the
kinematic distance, whose uncertainty is typically 20 \%,
 we expect a similar
uncertainty to exist in the expected rotational velocity. The expected
velocity shift is also dependent on the assumed inclination of the cloud.

}

\begin{figure*}
\includegraphics[width=1 \textwidth]{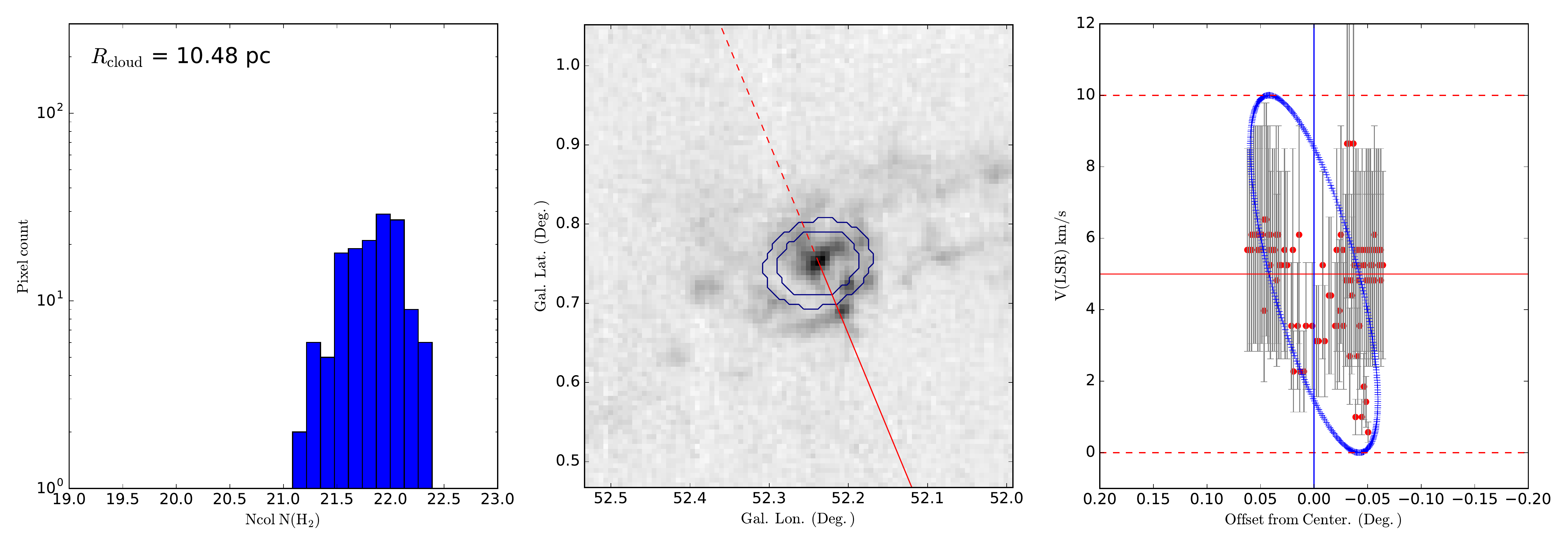}
\includegraphics[width=1 \textwidth]{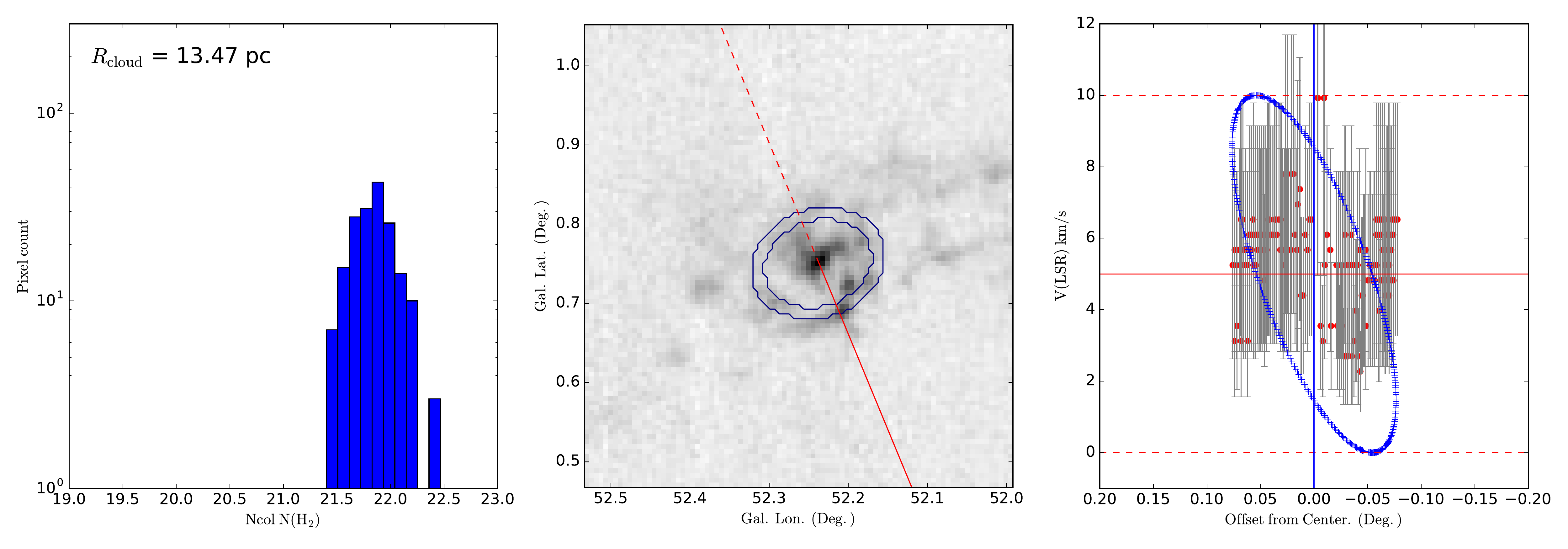}
\includegraphics[width=1 \textwidth]{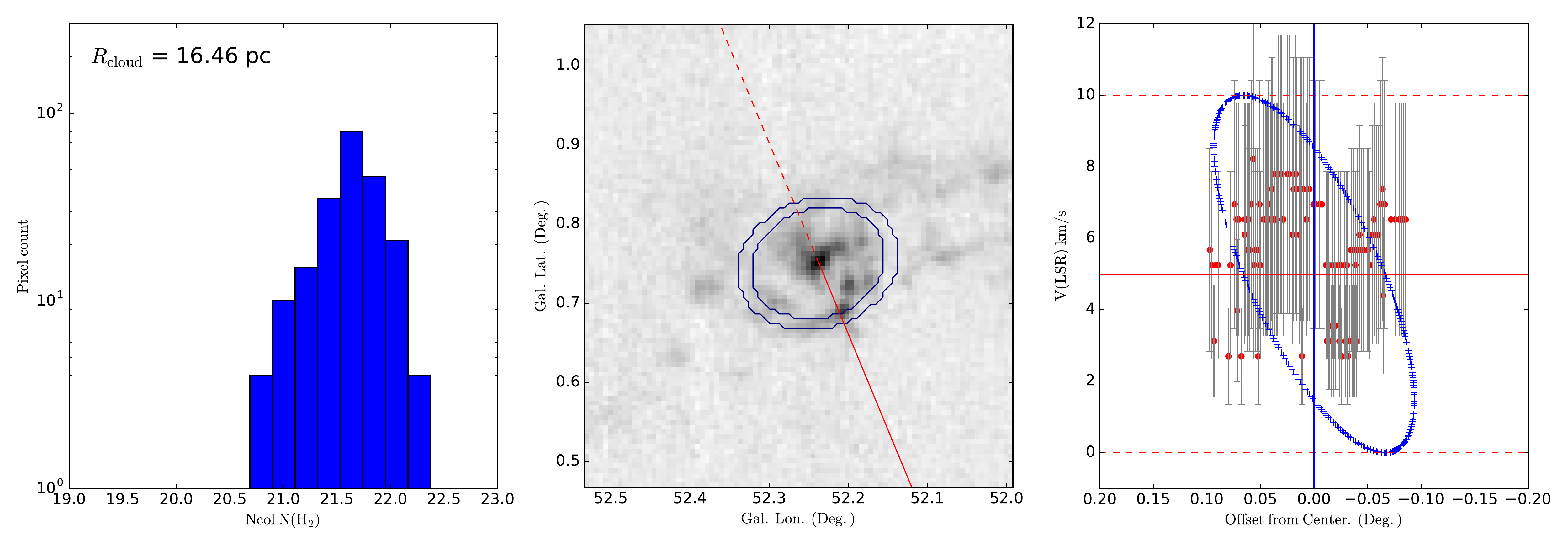}
\caption{A continuation of Fig. \ref{fig:rings1}. \label{fig:rings2}}
\end{figure*}
\begin{figure*}
\includegraphics[width=1 \textwidth]{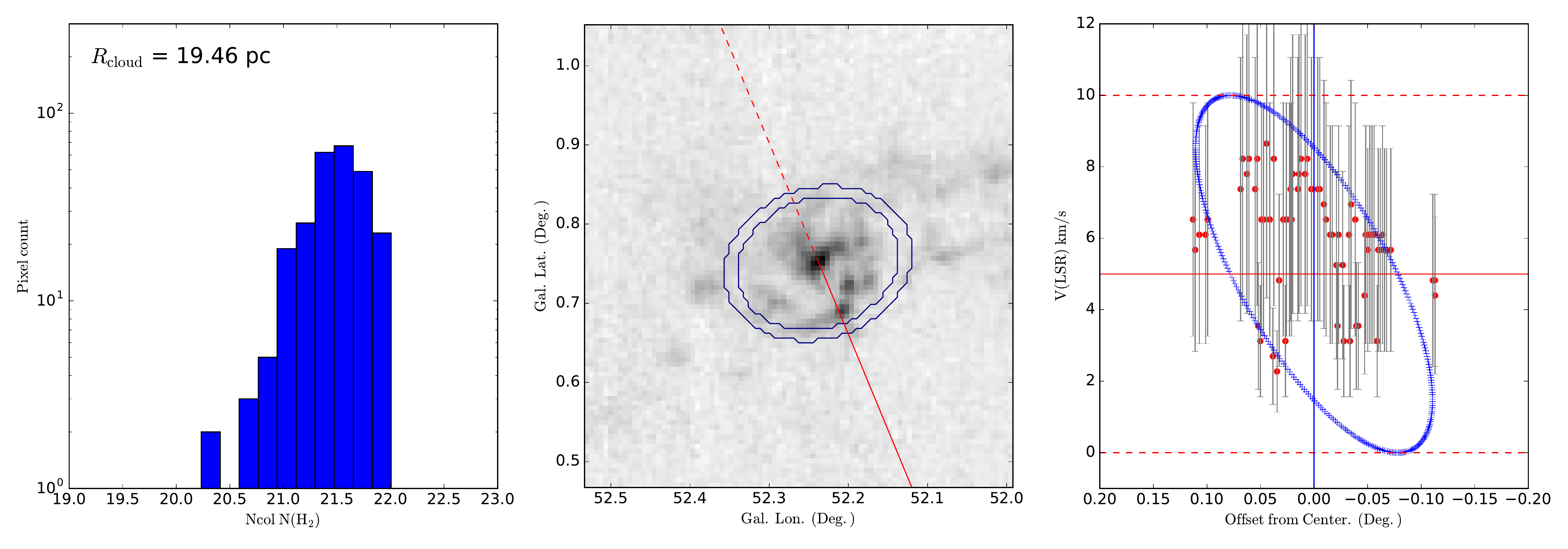}
\includegraphics[width=1 \textwidth]{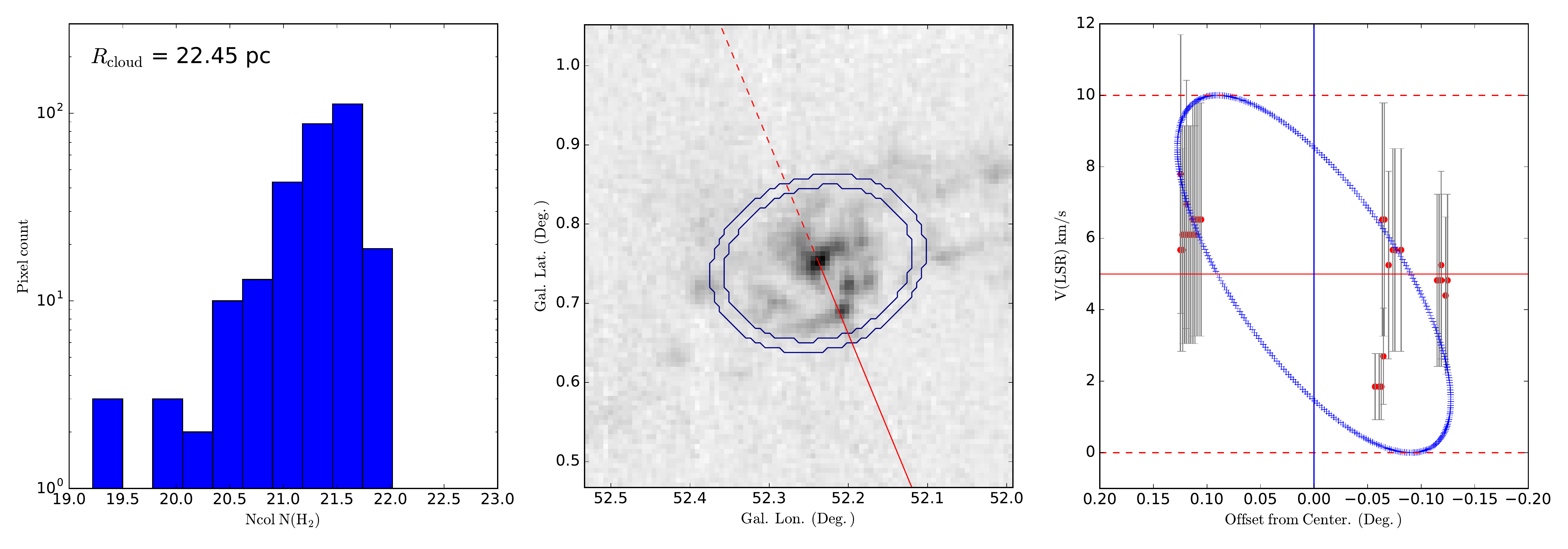}
\includegraphics[width=1 \textwidth]{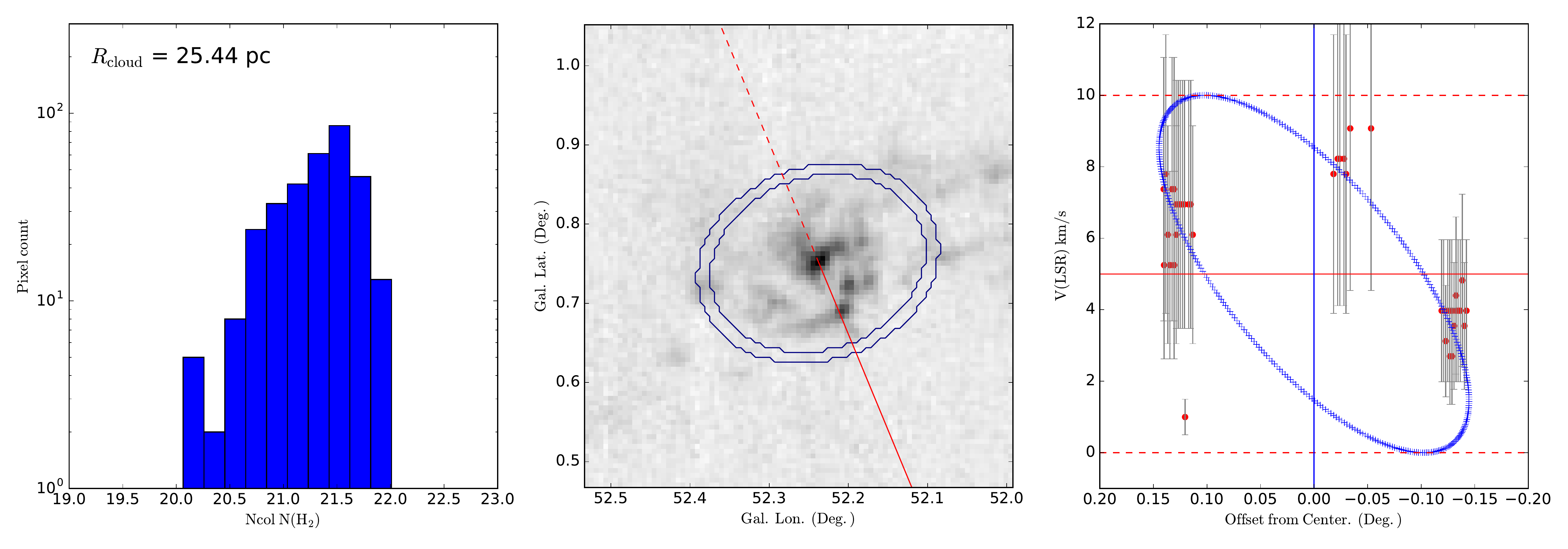}
\caption{A continuation of Fig. \ref{fig:rings1}, \ref{fig:rings2}.
\label{fig:rings3}}
\end{figure*}

\section{Detailed velocity structure of the spiral cloud}\label{sec:appen:v}
\begin{figure*}
\includegraphics[width = 1 \textwidth]{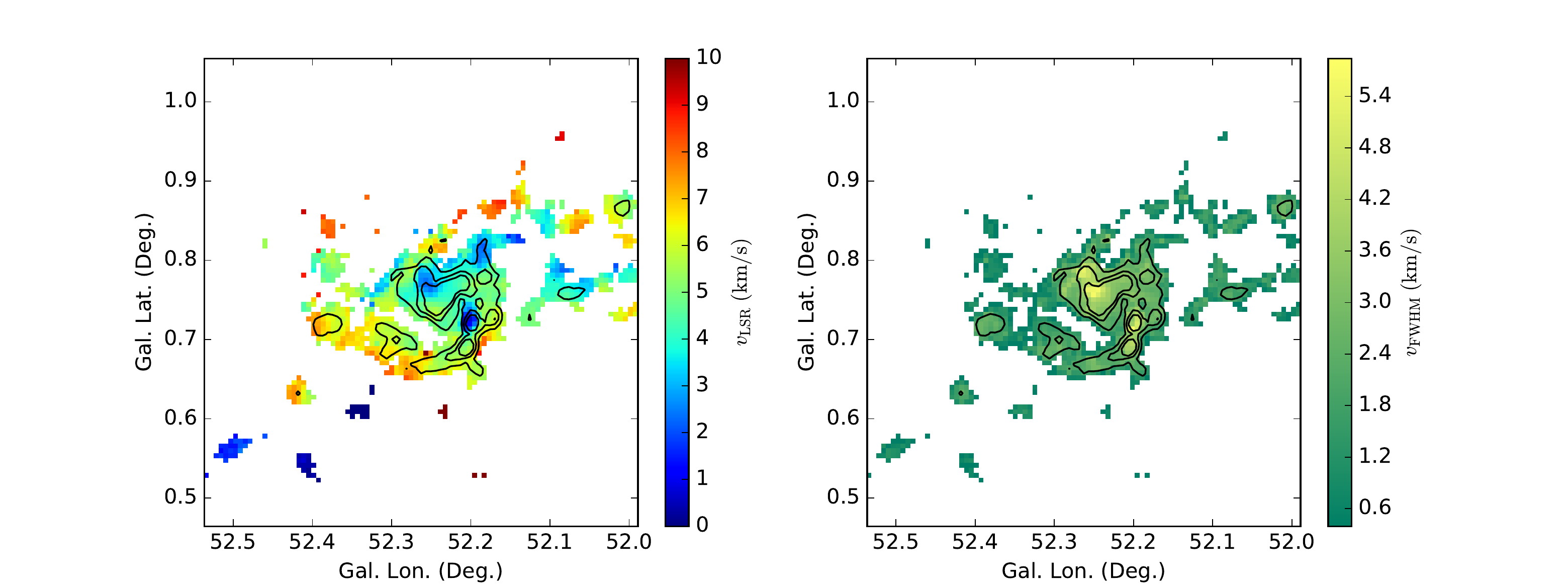}
\caption{  Detailed velocity structure of the spiral cloud. The left panel
shows the velocity centroid map, and the right panel shows the velocity FWHM
map.
Overlaid contours are velocity-integrated $^{13}$CO(1-0) emission, and the levels correspond to 5.1, 10.2, 15.3 $\rm K\; km\;s^{-1}$.
In producing these maps, we have excluded vorxels where the emission is below 0.43
K, which corresponds to three times the rms noise level of the
data cube.\label{fig:mom} }
\end{figure*}

In Fig. \ref{fig:mom} we present the velocity centroid map and the velocity FWHM
map of the cloud.

\end{appendix}

%\begin{figure*}
%\includegraphics[width=1.0 \textwidth]{co_int-eps-converted-to.pdf}
%\caption{Velocity-integrated $^{13}$CO emission. \label{fig:5152} }
%\end{figure*}

%\noindent 
%The cloud  belongs to cloud pair. The cloud G52.24 exhibit a clear spiral-shaped geometry, and the cloud G51.69 have a ``normal'' geometry. These two clouds seems to be connected by some filaments (Figure \ref{fig:5152}). 
%
%\subsection{Cloud Kinematics}
%The cloud G52.24 have a clear velocity gradient, indicative of rotation. Here we quantify the dynamical structure of the cloud by deriving the column density, profile,  and velocity dispersion profiles. Here, the column density of the cloud is estimated as
%\begin{equation}
%N(H_2)=4.92\times 10^{20}\times T_{mb}\times v\; ,
%\end{equation}
%
% 
%\begin{figure*}
%\includegraphics[width=0.4\textwidth]{dyn_cloud-eps-converted-to.pdf}
%\includegraphics[width=0.4\textwidth]{dyn_ncol-eps-converted-to.pdf}\\
%\caption{Kinematic Properties of the cloud. Upper left: structure of the cloud. We divide the cloud into different sections too derive its structure. Upper right. Column density structure of the cloud. Lower Left: Rotational Velocity. Lower Right: Velocity dispersion.}
%\end{figure*}

%
%\noindent
%{  Is the cloud rotationally supported?}
%\begin{equation}
%v_{\rm rot}\sim \sqrt{G\;\Sigma\;r}
%\end{equation}
%
\bibliography{paper}

\end{document}